\documentclass{iopjournal}

\usepackage[T1]{fontenc}
\IfFileExists{lmodern.sty}{\usepackage{lmodern}}{}

\usepackage{amsmath,amssymb,amsthm,bm,mathtools}
\usepackage{cite}
\usepackage{hyperref}
\renewcommand{\orcid}[1]{\href{https://orcid.org/#1}{\textnormal{\scriptsize [ORCID]}}}

\hypersetup{
	colorlinks=true,
	linkcolor=blue,
	citecolor=blue,
	urlcolor=blue,
	pdftitle={Geometric control of diffusive relaxation},
	pdfauthor={Pengfei Zhu}
}

\ifdefined\pdfminorversion
\fi

\renewcommand{\articletype}[1]{%
 \noindent{\large\sffamily Journal of Physics A: Mathematical and Theoretical}\par\vspace{6mm}%
 \noindent{\small\sffamily\MakeUppercase{#1}}\par\vspace{3mm}}

\newtheorem{proposition}{Proposition}[section]

\newcommand{\dd}{\mathrm d}
\newcommand{\ee}{\mathrm e}
\newcommand{\ii}{\mathrm i}
\newcommand{\cv}{c_{\mathrm v}}
\newcommand{\Dm}{D_{\mathrm{mac}}}
\newcommand{\av}[1]{\left\langle #1\right\rangle}
\newcommand{\Id}{\mathsf I}

\begin{document}

\articletype{Paper} 

\title{Geometric control of diffusive relaxation: spectral design, transport bounds and transverse mode isolation}

\author{Pengfei Zhu$^{1,*}$\orcid{0000-0003-0431-1538}}

\affil{$^1$Bundesanstalt für Materialforschung und -prüfung (BAM), 12205 Berlin, Germany}

\affil{$^*$Author to whom any correspondence should be addressed.}

\email{pengfei.zhu@bam.de}


\begin{abstract}
Geometric confinement changes both thermal relaxation and heat throughput,
but these effects need not be governed by the same conduction path. We
establish a distinction between serial and parallel geometries for passive,
homogeneous Fourier conductors at fixed material properties, volume and
length. For a reduced variable-area conductor, two logarithmic shape
harmonics independently tune two selected Bloch-rate pairs, and exact
local transport level sets permit spectral redistribution at fixed
effective diffusivity. A nonperturbative weighted Poincar\'e bound,
however, requires the effective axial diffusivity to vanish if the first
positive insulated rate tends to zero at fixed length. This obstruction
is avoided by continuous parallel paths with weak transverse exchange.
Capacity-weighted networks identify the exchange mechanism, and a
three-dimensional product domain provides an exact bulk realization.
Its axial conductance remains $\kappa V/L^2$, while exactly one positive
insulated rate vanishes and the next approaches a positive limit set by
axial, chamber and neck diffusion. The slow rate is asymptotically the
transverse neck conductance times the sum of inverse chamber capacities.
Spectral projection estimates translate this separation into a
preparation- and readout-dependent observation window. The result is
metastable retention of a temperature contrast under insulated conditions,
not a second equilibrium state. It identifies when geometrical control
of internal equilibration can be separated from macroscopic heat transport.
\end{abstract}

\begingroup
\hypersetup{linkcolor=black}
\renewcommand{\contentsname}{Contents}
\setcounter{tocdepth}{2} 
\tableofcontents
\endgroup

\section{Introduction}
Heat conduction is conventionally described by Fourier's law and the associated diffusion equation, which relate the relaxation of a temperature field to its spatial gradients and the thermal properties of the medium\cite{Zhu2026Exceptional,Zhu2026Snell,Zhu2026SpectralGeometry}. In a homogeneous conductor, the intrinsic diffusivity sets a natural rate scale, while the specimen dimensions and boundary conditions determine the decay of individual temperature patterns. For an insulated finite body, the spatially uniform temperature is a conserved mode. All nonuniform modes decay exponentially at rates given by the positive spectrum of a self-adjoint diffusion operator. This spectral viewpoint connects transient heat conduction to classical questions of dissipation, equilibration, and the approach to thermal equilibrium. It also distinguishes the decay of a particular temperature pattern from the steady throughput measured under an imposed temperature difference. Although both processes originate from the same local Fourier law, they probe different boundary-value problems and need not respond identically to geometric changes. \cite{Simoncelli2020,Gaspard2008,Garden2022}

The role of geometry in diffusion has a long history. Variable cross sections, constrictions, and confinement modify the balance between local storage and longitudinal flux, even when the material diffusivity is spatially uniform\cite{Holyst2019,Lamba2025,Haller2018}. Cross-sectionally averaged descriptions, including Fick--Jacobs-type reductions and diffusion in entropic landscapes, provide an effective framework for transport through corrugated channels\cite{Burada2007,Kalinay2011}. Related homogenization theories characterize the large-scale diffusivity of periodic structures\cite{Torquato2021,Papaioannou2017}, whereas spectral and variational methods characterize the decay of nonuniform fields in finite domains\cite{BuenoOrovio2006,Vay2013}. Narrow passages can produce widely separated time scales by permitting rapid equilibration within larger regions but only weak exchange between them. These observations motivate a distinction that is central to the present work: the effective transport coefficient describes the long-wavelength response, whereas the nonzero relaxation eigenvalues describe a hierarchy of internal decay channels.

Geometric manipulation of heat transfer has also become a principal theme of thermal metamaterials and thermal-device design. Transformation-based thermotics and effective-medium approaches have enabled the theoretical and experimental study of thermal cloaks\cite{Li2021}, concentrators\cite{Li2019Doublet}, rotators\cite{Sun2025}, and other structures that redistribute temperature gradients and steady conductive heat fluxes. Subsequent developments have extended the design space toward transient responses\cite{Li2026Electrothermal}, time-dependent material properties\cite{Yang2024}, and coupled transport processes\cite{Schittny2013}. In many of these settings, the primary performance measure is an externally prescribed temperature or flux field, an effective constitutive parameter, or the spatial distribution of heat flow. A complementary objective is to tailor the intrinsic decay spectrum of a passive body after the external preparation has ended. This requires asking not only where heat flows, but which temperature patterns disappear first, which remain long lived, and whether a desired slow pattern can coexist with useful macroscopic conduction.

An instructive comparison comes from wave-based spectral engineering. In electronic crystals\cite{Wang2015}, photonic crystals\cite{Perczel2017}, and phononic or acoustic metamaterials\cite{Aleksandravicius2026}, periodic structure couples spatial harmonics and lifts degeneracies at selected Bloch wave numbers. Bragg scattering, localized resonances, and network connectivity can thereby reshape dispersions and isolate selected wave modes. Periodic diffusion operators admit a related Bloch decomposition, but their eigenvalues have a fundamentally different interpretation: they are real, nonnegative relaxation rates rather than oscillation frequencies or propagation energies\cite{Zhu2026GVWT,Zhu2026Unified,Mandelis2001,Tabatabaei2011}. A gap between diffusive Bloch branches therefore represents a separation of exponential decay rates, not a forbidden propagation band. Moreover, reshaping selected Bloch branches does not automatically isolate the slowest nonconserved eigenmode of a finite body. The latter is a global spectral question involving the entire low-energy subspace and the specimen's boundary conditions.

Further connections arise in the spectral theory of reversible Markov processes\cite{Bovier2002,Jiang2003}, metastability\cite{Assaf2006}, and diffusion on weakly connected domains\cite{Cesbron2018}. For reversible diffusion, a small positive eigenvalue corresponds to a slowly relaxing component of the dynamics, while the separation between the first and subsequent positive eigenvalues controls whether a single long-lived component dominates after faster transients have decayed. Weighted Poincar\'e inequalities and variational principles relate such eigenvalues to geometric constraints and transport bottlenecks\cite{GarbunoInigo2020,Haller2020}. Thin-neck and dumbbell domains offer especially transparent examples: nearly disconnected regions support slow inter-region exchange, yet additional slow modes may emerge unless the rest of the spectrum is controlled. This perspective suggests a design principle based not merely on reducing the first relaxation rate, but on creating exactly one anomalously slow nonconserved mode while retaining a finite spectral scale for all competing modes.
These developments leave a specific question for passive Fourier conductors. Can one preserve axial heat throughput, at fixed material coefficients, specimen length, and total material volume, while making one internally stored temperature contrast arbitrarily long lived? The answer cannot be inferred from local band splitting alone, nor from the existence of a narrow thermal bottleneck. In a one-dimensional serial geometry, the same constriction that impedes exchange also impedes the steady axial current. In a multiply connected geometry, by contrast, axial transport and transverse exchange may occupy different spatial sectors. Establishing the distinction requires both a quantitative obstruction in the serial class and an explicit connected three-dimensional construction in which the slowest positive rate tends to zero but the next one remains bounded away from zero.

Here we develop a spectral framework for this problem in homogeneous Fourier conductors. We first study a capacity-weighted, variable-area diffusion operator and show that two harmonics of the logarithmic thickness provide independent local control of two selected Bloch relaxation-rate pairs. We retain the complete geometry-induced potential in the second-order calculation and identify local design paths along which the exact reduced macroscopic diffusivity is fixed. We then derive a sharp weighted Poincar\'e bound linking the first positive rate of a serial conductor to its effective axial transport, demonstrating the obstruction to arbitrarily slow relaxation at finite throughput within this class. To overcome it, we introduce capacity-weighted parallel conducting networks, in which longitudinal diffusion and weak transverse exchange contribute additively to the relaxation spectrum. Finally, an exactly separable, fixed-volume three-dimensional prism with a thin-necked transverse section establishes the bulk result: its steady axial conductance is identical to that of the reference prism, one positive insulated relaxation rate vanishes in the narrow-neck limit, and the next positive rate approaches a strictly positive scale. Spectral projection estimates then specify when the resulting temperature contrast can be prepared and observed. The resulting picture links geometric spectral design, transport bounds, and metastable thermal relaxation without changing the intrinsic material properties or invoking non-Fourier transport.

\section{Reversible diffusion and spectral design}
\label{sec:framework}

\subsection{Conservation, dissipation and the relaxation operator}

Let $\Omega\subset\mathbb R^3$ be a bounded, connected Lipschitz domain
occupied by a homogeneous Fourier conductor of volume $V$, conductivity
$\kappa>0$ and volumetric heat capacity $\cv>0$.
Both material coefficients remain fixed when the geometry changes,
and $D_0=\kappa/\cv$ denotes the intrinsic diffusivity. After preparation,
the temperature deviation $\theta=T-T_{\mathrm{eq}}$ obeys
\begin{equation}
	\cv\partial_t\theta=\kappa\Delta\theta,
	\qquad \partial_n\theta=0\quad\text{on }\partial\Omega,
	\qquad \int_\Omega\cv\theta\,\dd V=0.
	\label{eq:bulkheat}
\end{equation}
The final equilibrium temperature is spatially uniform. The constraint in
\eqref{eq:bulkheat} subtracts conserved heat and leaves the nonconserved
relaxation dynamics.

Introduce the capacity inner product and the conductive Dirichlet form,
\begin{equation}
	(u,v)_C=\int_\Omega\cv u^*v\,\dd V,
	\qquad \mathfrak a(u,v)=\int_\Omega\kappa\nabla u^*\cdot\nabla v\,\dd V.
	\label{eq:forms}
\end{equation}
The nonnegative self-adjoint rate operator $\mathcal L$ is defined weakly
by $(u,\mathcal Lv)_C=\mathfrak a(u,v)$, with the insulated boundary
conditions included in its domain. Thus
\begin{equation}
	\partial_t\theta=-\mathcal L\theta,
	\qquad
	\frac{\dd}{\dd t}\frac{\|\theta\|_C^2}{2}
	=-\mathfrak a(\theta,\theta)\leq0.
	\label{eq:dissipation}
\end{equation}
The quadratic measure $\|\theta\|_C^2/2$ is stored temperature
nonuniformity, rather than net excess heat. For small deviations from
$T_{\mathrm{eq}}>0$, the thermodynamic entropy deficit is
$\|\theta\|_C^2/(2T_{\mathrm{eq}}^2)$ to quadratic order.
More explicitly, for positive absolute temperature and constant $\cv$,
the entropy difference at conserved total energy is
\begin{equation}
 S_{\mathrm{eq}}-S[T]
 =\int_\Omega\cv\ln\frac{T_{\mathrm{eq}}}{T}\,\dd V
 =\frac{\|\theta\|_C^2}{2T_{\mathrm{eq}}^2}
 +O\!\left(\int_\Omega\frac{\cv|\theta|^3}{T_{\mathrm{eq}}^3}\,\dd V\right),
 \qquad
 \dot S=\int_\Omega\frac{\kappa|\nabla T|^2}{T^2}\,\dd V\geq0.
 \label{eq:entropyidentity}
\end{equation}
Thus a slow positive rate describes slow dissipation of spatial
nonuniformity after the conserved mean has been removed.

The eigenvalue problem and its Rayleigh quotient are
\begin{equation}
	\mathfrak a(v,u_n)=\lambda_n(v,u_n)_C,
	\qquad
	\lambda_n=\frac{\mathfrak a(u_n,u_n)}{\|u_n\|_C^2},
	\qquad
	0=\lambda_0<\lambda_1\leq\lambda_2\leq\cdots.
	\label{eq:generalrayleigh}
\end{equation}
The zero mode is constant; every positive-rate mode has zero
capacity-weighted mean. We distinguish two spectral objectives. Local
spectral design adjusts selected eigenvalue branches by small shape
changes. Strong mode isolation instead requires a family for which
$\lambda_1\to0$ while $\liminf\lambda_2>0$, at fixed macroscopic dimensions.
Splitting a pair of Bloch rates does not by itself establish the second
property.
The compact embedding $H^1(\Omega)\hookrightarrow L^2(\Omega)$
gives a discrete spectrum. In the mean-zero space
$H^1_*:=\{u\in H^1(\Omega):(1,u)_C=0\}$, the min--max formula reads
\begin{equation}
 \lambda_n=\min_{\substack{E\subset H^1_*\\\dim E=n}}
 \ \max_{u\in E\setminus\{0\}}
 \frac{\mathfrak a(u,u)}{\|u\|_C^2},\qquad n\geq1.
 \label{eq:minmax}
\end{equation}
A single trial function can therefore show that $\lambda_1$ is small;
excluding a second slow mode requires control of the rest of the
low-energy space.

\subsection{Throughput and relaxation as distinct boundary-value problems}

Axial throughput is measured in a separate experiment: impose constant
temperatures on the complete terminal faces at $x=0,L$ and insulate the
lateral boundary. Define the steady conductance and its fixed-volume
reference by
\begin{equation}
	\mathcal G_{\mathrm{ss},D}=\frac{|Q_{\mathrm{ss}}|}{|T_R-T_L|},
	\qquad
	\mathcal G_{\mathrm{ref}}=\frac{\kappa V}{L^2}.
	\label{eq:conductanceprotocol}
\end{equation}
The reference is a prism of identical material, volume and length.
Superscripts $N$ and $D$ below distinguish fully insulated relaxation
from terminally contacted conduction. Comparing these two measurements
asks whether one geometry can support efficient axial transport and a
slow insulated internal contrast. It does not equate their spectra.

\subsection{A positive one-dimensional profile and its invariant measure}

For a strip of fixed width $b$, periodic thickness $h(x+a)=h(x)>0$ and
$L=Na$, cross-sectional equilibration gives the leading balance
\begin{equation}
	\cv b h\,\partial_t\theta=\partial_x(\kappa b h\partial_x\theta),
	\qquad
	\mathcal L_h=-D_0h^{-1}\partial_x(h\partial_x).
	\label{eq:thinheat}
\end{equation}
The same area factor contributes to storage and conduction. This
Fick--Jacobs form is valid when the retained modal rates are small compared
with transverse equilibration rates and the wall slope is small
\cite{Zwanzig1992,Reguera2001,Kalinay2005,Kalinay2006,Kalinay2026}.
All formulas derived from \eqref{eq:thinheat} are exact for that reduced
operator; they are not unrestricted identities for an arbitrary thick
three-dimensional strip. In particular, no finite-slope correction is
used as an exact closure.
Uniformity across the fixed-width direction is either imposed as an
invariant preparation sector or requires the retained rates to be small
compared with $D_0\pi^2/b^2$. This additional scale matters when the
reduced spectrum is compared with the full three-dimensional spectrum.

Fix $a=2\pi/q$ and the mean thickness $\bar h$. A convenient design space is
\begin{equation}
	s=qx,\qquad
	h(x;\bm\alpha)=\bar h\,\frac{\ee^{2W(s;\bm\alpha)}}{Z(\bm\alpha)},
	\qquad
	Z=\av{\ee^{2W}}_s,\qquad
	\av{f}_s=\frac1{2\pi}\int_0^{2\pi}f(s)\,\dd s.
	\label{eq:normalizedh}
\end{equation}
Then $V=bL\bar h$ and the total heat capacity remain fixed. Multiplying
$h$ by a spatial constant cancels from \eqref{eq:thinheat}; normalization
preserves the material budget without changing the reduced spectrum.

The backward diffusion generator $-\mathcal L_h$ is reversible in the
measure $\pi(x)\dd x=h(x)\dd x/M$, where $M=\int_0^a h\,\dd x$.
Its forward equation for a probability density $p$ is
\begin{equation}
	\partial_t p=D_0\partial_x\!\left[h\partial_x(p/h)\right],
	\qquad p_{\mathrm{eq}}=\frac hM.
	\label{eq:forward}
\end{equation}
This establishes the connection with diffusion in an entropic landscape.
The stationary density proportional to $h$ describes cross-sectional
weighting; it does not imply a nonuniform equilibrium temperature.
Indeed, putting $p=h\theta/M$ maps the backward temperature equation
to the forward density equation. This is a probability interpretation
only when the initial $\theta$ is nonnegative and normalized; general
mean-zero thermal contrasts instead correspond to signed fields.

\section{Local control of the Bloch relaxation spectrum}
\label{sec:localcontrol}

\subsection{Sensitivity to geometric redistribution}

For fixed periodic, Bloch or insulated end conditions, normalize a simple
eigenfunction of \eqref{eq:thinheat} by $\int h|u_n|^2\dd x=1$.
Differentiating its generalized eigenvalue problem gives
\begin{equation}
	\delta\lambda_n=\int\delta h\,
	\bigl(D_0|u_n'|^2-\lambda_n|u_n|^2\bigr)\,\dd x.
	\label{eq:shapesensitivity}
\end{equation}
The variation of the eigenfunction cancels by stationarity of the
Rayleigh quotient. Adding material at a large temperature amplitude
increases storage, whereas adding material at a large gradient increases
dissipation. Their difference supplies the relevant mode-specific
sensitivity; geometric control cannot be inferred from local heat
capacity alone.

In the normalized family \eqref{eq:normalizedh},
$\delta\ln h=2(\delta W-\av{\delta W}_h)$, with
$\av{f}_h=\int hf\,\dd x/\int h\,\dd x$. The constant subtraction makes
no contribution because the integrated sensitivity vanishes. Hence
\begin{equation}
	J_{j\mu}:=\frac{\partial\lambda_{n_j}}{\partial\alpha_\mu}
	=2\int h\frac{\partial W}{\partial\alpha_\mu}
	\bigl(D_0|u_{n_j}'|^2-\lambda_{n_j}|u_{n_j}|^2\bigr)\,\dd x.
	\label{eq:spectraljacobian}
\end{equation}
For $m$ selected simple branches, full row rank of $J$ is the local
controllability condition for $m$ nearby spectral targets. A constraint
on another rate restricts admissible shape changes to the corresponding
Jacobian nullspace. At a degeneracy, \eqref{eq:shapesensitivity} is a
Hermitian matrix on the degenerate eigenspace and must first be
diagonalized there. Counting geometric parameters without this rank
condition can overestimate the number of independent spectral controls.

\subsection{The complete transformed operator}

The unitary change of weight, up to a constant factor,
$u=\ee^{-W}\psi$ gives
\begin{equation}
	\widehat{\mathcal L}=D_0q^2H,
	\qquad
	H=-\partial_s^2+W''+(W')^2
	=(-\partial_s-W')(\partial_s-W').
	\label{eq:schrodinger}
\end{equation}
Primes in this subsection denote $s$ derivatives. The factorization
preserves nonnegativity and the transformed zero mode $\psi_0\propto\ee^W$.
The effective potential is fixed by geometry; it is not an independently
specified material interaction. The evolution remains a sum of real
exponential decays.
The operator domain must be transformed together with the differential
expression. Bloch conditions are preserved by periodic $W$, whereas
an insulated endpoint $u'=0$ becomes
$(\partial_s-W')\psi=0$. These transformed conditions are also the
ones for which the factorization in \eqref{eq:schrodinger} is
nonnegative. In a Bloch sector the zero mode exists only at $k=0$.

Take the two-harmonic logarithmic profile
\begin{equation}
	W(s)=\eta\cos s+\xi\cos(2s+\phi).
	\label{eq:twoharmonics}
\end{equation}
For Bloch number $K=qk$, $-1/2\leq k\leq1/2$, expand
$\psi=\sum_m c_m\ee^{\ii(m+k)s}$. Then
\begin{equation}
	H_{mn}(k)=(m+k)^2\delta_{mn}+U_{m-n},
	\qquad
	U_\ell=-\ell^2W_\ell-\sum_r r(\ell-r)W_rW_{\ell-r},
	\label{eq:fullbloch}
\end{equation}
where $W_\ell$ and $U_\ell$ are Fourier coefficients. For
\eqref{eq:twoharmonics} the nonzero coefficients are exactly
\begin{align}
	U_0&=\eta^2/2+2\xi^2, &
	U_1&=-\eta/2+\eta\xi\ee^{\ii\phi},\nonumber\\
	U_2&=-2\xi\ee^{\ii\phi}-\eta^2/4, &
	U_3&=-\eta\xi\ee^{\ii\phi},\nonumber\\
	U_4&=-\xi^2\ee^{2\ii\phi}, & U_{-\ell}&=U_\ell^*.
	\label{eq:fouriercoefficients}
\end{align}
Two shape harmonics therefore generate four operator harmonics. Retaining
only a direct coupling coefficient can give an incorrect splitting.
The term ``Bragg splitting'' here refers to the lifting of a degeneracy
between spatial harmonics whose wave numbers differ by a reciprocal
lattice vector. Its physical consequence is a difference of decay
times and spatial patterns. Since the eigenvalues are relaxation rates,
the splitting does not by itself describe a propagating-wave stop band.

\subsection{Two independently adjustable pairs}

\begin{figure}[!htbp]
\centering
\includegraphics[width=0.94\linewidth]{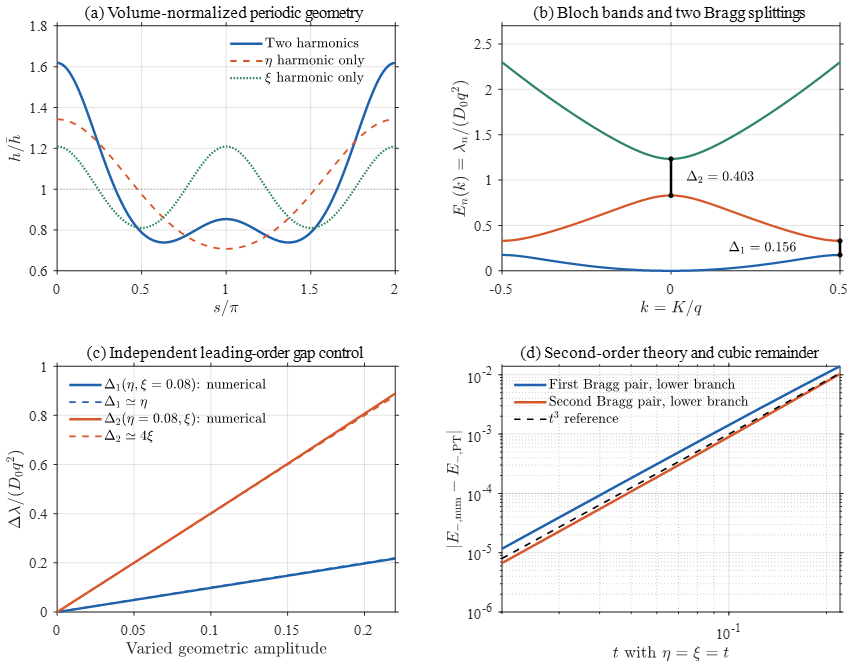}
\caption{\textbf{Local control of two Bloch relaxation pairs.}
(a) Fixed-mean thickness profiles with both logarithmic harmonics and
with either harmonic separately. (b) The lowest Bloch-rate branches;
the marked separations occur at the first zone edge and at the second
free degeneracy at zero Bloch number. (c) Pair splittings from the
Fourier operator compared with the leading laws
$\Delta\lambda_1/(D_0q^2)=\eta$ and
$\Delta\lambda_2/(D_0q^2)=4\xi$. (d) Lower-branch errors of the
second-order formulas along $\eta=\xi=t$, compared with a cubic
reference. The split branches describe exponential relaxation of
spatial patterns.}
\label{fig:fig1}
\end{figure}

Let $\rho=|\eta|+|\xi|\ll1$. For branch derivatives take $\eta,\xi>0$
of comparable order, away from crossings. The first free degeneracy
occurs at $k=1/2$ between momenta $\pm1/2$; the second occurs at $k=0$
between momenta $\pm1$. Eliminating remote momenta through second order
gives, in the positive/negative-momentum bases,
\begin{align}
	H_{1,\mathrm{eff}}&=
	\left(\frac14+\frac{3\eta^2}{8}-\frac{2\xi^2}{3}\right)\Id
	+\begin{pmatrix}0&-\eta/2\\-\eta/2&0\end{pmatrix}
	+O(\rho^3),\label{eq:effectivepair1}\\
	H_{2,\mathrm{eff}}&=
	\left(1+\frac{2\eta^2}{3}+\frac{3\xi^2}{2}\right)\Id
	+\begin{pmatrix}0&-2\xi\ee^{\ii\phi}\\-2\xi\ee^{-\ii\phi}&0\end{pmatrix}
	+O(\rho^3).
	\label{eq:effectivepair2}
\end{align}
The nominal second-order off-diagonal terms cancel against
intermediate-state contributions, as derived in
Appendix~\ref{app:perturbation}. The rates are
\begin{align}
	\frac{\lambda_{1,\pm}}{D_0q^2}
	&=\frac14+\frac{3\eta^2}{8}-\frac{2\xi^2}{3}
	\pm\frac\eta2+O(\rho^3),\label{eq:blochrates1}\\
	\frac{\lambda_{2,\pm}}{D_0q^2}
	&=1+\frac{2\eta^2}{3}+\frac{3\xi^2}{2}
	\pm2\xi+O(\rho^3).
	\label{eq:blochrates2}
\end{align}
The pair index here labels a free Bloch degeneracy, not the ordering of
all eigenvalues of a finite specimen. In particular,
\begin{equation}
	\Delta\lambda_1=D_0q^2[\eta+O(\rho^3)],\qquad
	\Delta\lambda_2=D_0q^2[4\xi+O(\rho^3)].
	\label{eq:twogaps}
\end{equation}
The lower-branch Jacobian is
\begin{equation}
	\frac{\partial(\lambda_{1,-},\lambda_{2,-})}{\partial(\eta,\xi)}
	=D_0q^2\left[
	\begin{pmatrix}-1/2&0\\0&-2\end{pmatrix}+O(\rho)\right],
	\label{eq:rankexample}
\end{equation}
and has nonzero determinant for sufficiently small positive amplitudes.
These two rates can therefore be set locally and independently at fixed
volume and period (Fig.~\ref{fig:fig1}). Their partners have opposite leading shifts about
their respective pair centres, so four rates have not become four
independent controls.

The slow transformed patterns are $\cos(s/2)$ and
$\cos(s+\phi/2)$ to leading order; the temperature patterns include the
factor $\ee^{-W}$. On a specimen with periodic ends, the first requires
an even number of cells. The phase $\phi$ shifts the second pattern but
does not affect either pair of rates through second order. It is
therefore not a third leading-order rate control.

\subsection{Macroscopic diffusion and a constrained spectral change}

\begin{figure}[!htbp]
\centering
\includegraphics[width=0.94\linewidth]{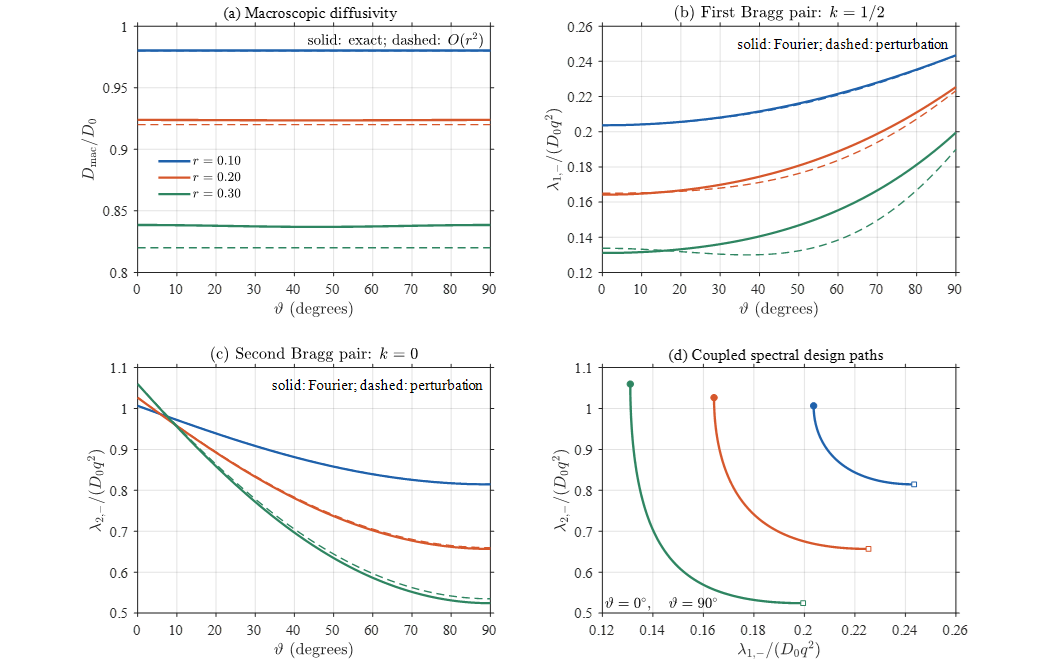}
\caption{\textbf{Redistributing relaxation rates at nearly fixed transport.}
The amplitudes are parametrized by $\eta=r\cos\vartheta$ and
$\xi=r\sin\vartheta$ for the three indicated values of $r$.
(a) Exact reduced effective diffusivity and its quadratic approximation.
(b,c) Lower rates of the first and second Bloch pairs, comparing Fourier
calculations with second-order perturbation theory. (d) The corresponding
rate trade-off curves. Fixed $r$ preserves transport through second
order; the exact diffusivity retains higher-order angular dependence.
The local exact transport level sets are described in
Sec.~\ref{sec:exacttransportlevel}.}
\label{fig:fig2}
\end{figure}

The long-wave branch has
$\lambda_0(K)=\Dm K^2+O(K^4)$, where the effective diffusivity is
\begin{equation}
	\Dm=\frac{D_0}{\av{h}_s\av{h^{-1}}_s}
	=\frac{D_0}{\av{\ee^{2W}}_s\av{\ee^{-2W}}_s}.
	\label{eq:lifsonjackson}
\end{equation}
This is the reversible periodic-diffusion formula
\cite{Lifson1962}; it also follows directly by averaging the series
thermal resistance and capacity.
The periodic corrector $\chi$ supplies a direct derivation:
\begin{equation}
 \partial_x[h(1+\chi')]=0,\qquad \chi(x+a)=\chi(x),\qquad
 h(1+\chi')=\frac{a}{\int_0^a h^{-1}\,\dd x},\qquad
 \Dm=D_0\frac{\av{h(1+\chi')^2}}{\av h}.
 \label{eq:cellcorrector}
\end{equation}
The last expression is the minimum dissipation for a unit mean
gradient, divided by mean storage, and gives
\eqref{eq:lifsonjackson}. For \eqref{eq:twoharmonics},
\begin{equation}
	\frac{\Dm}{D_0}=1-2(\eta^2+\xi^2)+O(\rho^4).
	\label{eq:macroweak}
\end{equation}
The product of the two averages is invariant under $W\mapsto-W$, so
odd total orders cancel. Setting $\eta=r\cos\vartheta$ and
$\xi=r\sin\vartheta$, with $r\ll1$ fixed, preserves $\Dm$ through
second order while
\begin{equation}
	\frac{\lambda_{1,-}}{D_0q^2}=\frac14-\frac r2\cos\vartheta+O(r^2),
	\qquad
	\frac{\lambda_{2,-}}{D_0q^2}=1-2r\sin\vartheta+O(r^2).
	\label{eq:redistribution}
\end{equation}
Figure~\ref{fig:fig2} illustrates this fixed-$r$ trade-off.
Thus changing the relative harmonic weights can prolong one selected
pattern and accelerate the other without a leading change of long-wave
transport. This is a perturbative redistribution of relaxation rates.
Whether an arbitrarily long lifetime can coexist with finite throughput
requires a separate, nonperturbative analysis.

\subsection{Redistribution along an exact transport level set}
\label{sec:exacttransportlevel}

The fixed-$r$ construction can be strengthened within the reduced
operator. Writing $m_j=\av{W^j}_s$ gives
\begin{equation}
 \av{\ee^{2W}}_s\av{\ee^{-2W}}_s
 =1+4m_2+4m_2^2+\frac43m_4+O(\rho^6).
 \label{eq:transportmoments}
\end{equation}
For \eqref{eq:twoharmonics},
$m_2=(\eta^2+\xi^2)/2$ and
$m_4=3(\eta^4+\xi^4)/8+3\eta^2\xi^2/2$, independently of $\phi$.
Inverting \eqref{eq:transportmoments} yields
\begin{equation}
 d(\eta,\xi):=\frac{\Dm}{D_0}
 =1-2(\eta^2+\xi^2)
 +\frac52(\eta^4+\xi^4)+4\eta^2\xi^2+O(\rho^6).
 \label{eq:macroquartic}
\end{equation}
At any sufficiently small point with $\eta,\xi>0$ of comparable order,
$\nabla d=-4(\eta,\xi)+O(\rho^3)\ne0$. The implicit-function theorem
therefore supplies a local curve on which the \emph{exact} value of
\eqref{eq:lifsonjackson} is constant. Along this curve,
\begin{equation}
 \frac{\dd\xi}{\dd\eta}=-\frac\eta\xi+O(\rho^2),\qquad
 \frac{\dd\lambda_{1,-}}{\dd\eta}
 =D_0q^2[-\tfrac12+O(\rho)],\qquad
 \frac{\dd\lambda_{2,-}}{\dd\eta}
 =D_0q^2[2\eta/\xi+O(\rho)].
 \label{eq:exactleveltangent}
\end{equation}
Increasing the first harmonic can thus slow the first selected pattern
and speed up the second while keeping the reduced steady conductance
exactly fixed. Equivalently, at a prescribed $d_*=1-2r_0^2$ close to
unity, an exact level curve has $r(\vartheta)=r_0+O(r_0^3)$ on compact
subintervals of $0<\vartheta<\pi/2$. This is local spectral control,
and does not assert an unlimited lifetime at fixed throughput.
With $\phi$ fixed, the two-amplitude design space has only one free
parameter after imposing $d=d_*$. Thus the two rates vary in a coupled
way along this level set: their unconstrained local independence in
\eqref{eq:rankexample} must not be interpreted as two independently
prescribable targets at fixed transport.

\section{A transport bound for serial diffusion}
\label{sec:serial}

\subsection{A weighted Poincar\'e bound expressed through conductance}

The one-dimensional model has only one axial conduction path. Its
capacity and resistance weights on a period are
\begin{equation}
	M=\int_0^a h\,\dd x,\qquad
	R=\int_0^a h^{-1}\,\dd x,\qquad
	\Dm=\frac{D_0a^2}{MR}.
	\label{eq:MR}
\end{equation}
The thermal resistance is $R/(\kappa b)$. On a specimen of $N$ repeated
cells, \eqref{eq:conductanceprotocol} and the series law give
\begin{equation}
	\mathcal G_{\mathrm{ss},D}^{(1\mathrm d)}=\frac{\kappa b}{NR},
	\qquad
	\frac{\mathcal G_{\mathrm{ss},D}^{(1\mathrm d)}}{\mathcal G_{\mathrm{ref}}}
	=\frac{a^2}{MR}=\frac{\Dm}{D_0}.
	\label{eq:seriesconductance}
\end{equation}
This identity pertains to the reduced variable-area conductor with
$V=NbM$; it makes the transport penalty a fixed-volume comparison.

\begin{proposition}[Serial relaxation--transport bound]
	\label{prop:serial}
	Let $h$ be a positive periodic profile bounded above and below on
	$[0,a]$. For the first positive cell-periodic rate of
	\eqref{eq:thinheat},
	\begin{equation}
		\lambda_1^{\mathrm{per}}\geq\frac{16D_0}{MR}
			=\frac{16\Dm}{a^2}.
		\label{eq:serialperiodic}
	\end{equation}
	For insulated ends on $[0,L]$, with $L=Na$,
	\begin{equation}
		\lambda_1^{N}\geq\frac{4\Dm}{L^2}.
		\label{eq:serialneumann}
	\end{equation}
	Consequently, for either choice of ends,
	\begin{equation}
		\frac{\tau_1}{\tau_{1,\mathrm{flat}}}
		\leq\frac{\pi^2}{4}\frac{D_0}{\Dm}
		=\frac{\pi^2}{4}
		\frac{\mathcal G_{\mathrm{ref}}}{\mathcal G_{\mathrm{ss},D}^{(1\mathrm d)}}.
		\label{eq:lifetimethroughput}
	\end{equation}
	Here $\tau_{1,\mathrm{flat}}=a^2/(4\pi^2D_0)$ for the cell-periodic
	problem and $L^2/(\pi^2D_0)$ for the insulated specimen.
\end{proposition}

\begin{proof}
	It suffices to consider real, nonconstant $u$ with $\int hu\,\dd x=0$.
	Let $d=u_{\max}-u_{\min}$. Weighted variance is bounded by
	\begin{equation}
		\int_0^a hu^2\,\dd x\leq\frac{Md^2}{4}.
		\label{eq:variancebound}
	\end{equation}
	For completeness, this follows by averaging
	$(u-u_{\min})(u_{\max}-u)\geq0$ and then using
	$-u_{\min}u_{\max}\leq d^2/4$ at zero mean. On the periodic interval,
	the maximum and minimum are joined by two arcs of resistance weights
	$R_1,R_2$, with $R_1+R_2=R$. Cauchy--Schwarz on each arc gives
	\begin{equation}
		\int_0^a h(u')^2\,\dd x
		\geq d^2\left(R_1^{-1}+R_2^{-1}\right)
		\geq\frac{4d^2}{R}.
		\label{eq:arcbound}
	\end{equation}
	Dividing by \eqref{eq:variancebound} proves
	\eqref{eq:serialperiodic}. With insulated ends, a single path between
	the extrema has resistance at most $NR$; its dissipation is at least
	$d^2/(NR)$, while the variance is at most $NMd^2/4$. This gives
	\eqref{eq:serialneumann}. Comparison with the flat rates proves
	\eqref{eq:lifetimethroughput}.
\end{proof}

\begin{figure}[!htbp]
\centering
\includegraphics[width=0.94\linewidth]{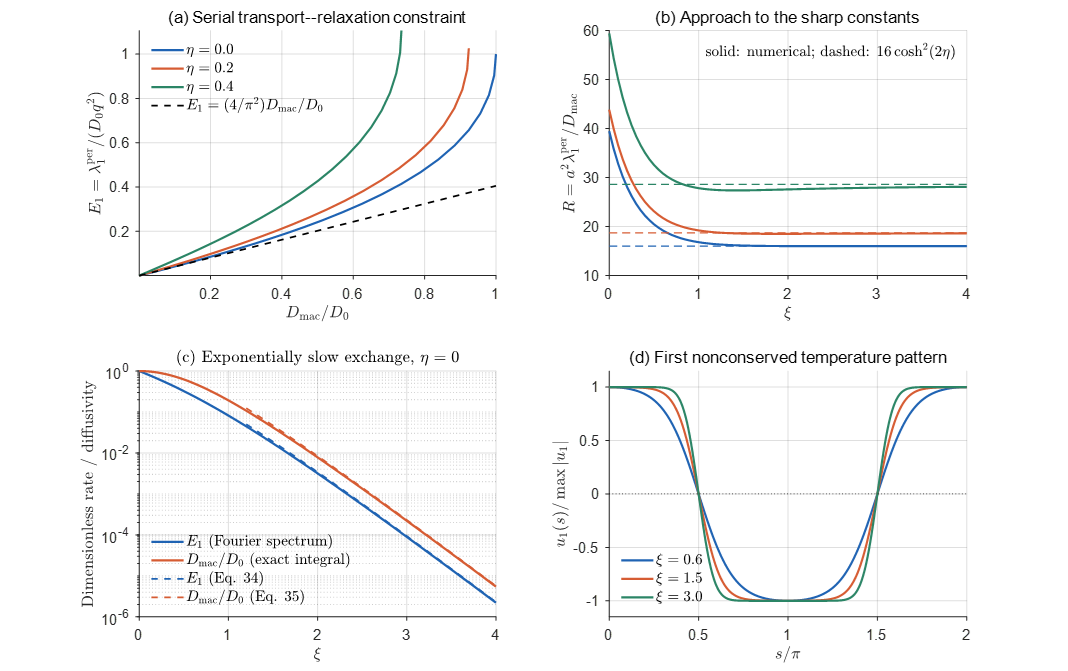}
\caption{\textbf{The serial relaxation--transport constraint.}
(a) First positive cell-periodic rate against reduced effective
diffusivity, with the lower bound in \eqref{eq:serialperiodic}.
(b) The dimensionless ratio $a^2\lambda_1^{\mathrm{per}}/\Dm$
approaches $16\cosh^2(2\eta)$ as the second harmonic strengthens.
(c) In the symmetric family, the rate and diffusivity both become
exponentially small, compared with the positive-integral asymptotics
\eqref{eq:strongrate} and \eqref{eq:strongmacro}.
(d) The slow temperature eigenfunction approaches a two-storage
contrast, with its main gradient in the resistive passages.
The sharp periodic constant is approached at $\eta=0$.}
\label{fig:fig3}
\end{figure}

Figure~\ref{fig:fig3} illustrates the bound and its sharp high-contrast limit.
The lower bound is uniform over profile contrast. A sequence whose
first positive rate tends to zero at fixed length must therefore have
$\Dm\to0$ in this serial class. It does not prohibit a large ratio of
arbitrary high-order rates, and it does not apply to a body with several
axial paths. The constants concern the specified one-dimensional
operators, rather than every Neumann geometry of the same volume.

\subsection{Strong modulation and the sharp periodic constant}

The high-contrast limit of \eqref{eq:twoharmonics} makes the physical
content of the bound explicit. Set $\phi=0$, let $\xi\to\infty$ and
hold $\eta$ bounded. In a cell there are two storage peaks near
$s=0,\pi$, joined by two narrow passages. For the cell-periodic sector,
let $C_A,C_B$ be the storage capacities and let $G$ be the conductance
of either passage. Fast equilibration within each storage region gives
\begin{equation}
	C_A\dot\Theta_A=-2G(\Theta_A-\Theta_B),\qquad
	C_B\dot\Theta_B=2G(\Theta_A-\Theta_B),
	\label{eq:serialreservoirs}
\end{equation}
where
\begin{equation}
	C_i=\cv b\int_{\mathcal A_i}h\,\dd x,\qquad
	G^{-1}=\int_{A\to B}\frac{\dd x}{\kappa b h},\qquad
	\lambda_{\mathrm{ex}}=2G(C_A^{-1}+C_B^{-1}).
	\label{eq:serialexchange}
\end{equation}
The factor two counts the two periodic passages. The exchange pattern
satisfies $C_A\Theta_A+C_B\Theta_B=0$.

Positive storage and resistance integrals, evaluated in
Appendix~\ref{app:asymptotics}, yield the reduced-operator asymptotics
\begin{align}
	C_{A,B}&=\frac{\pi\cv b\bar h}{q}
	\frac{\ee^{\pm2\eta}}{\cosh(2\eta)}[1+O(\xi^{-1})],\nonumber\\
	G&=\frac{4\kappa bq\bar h\xi}{\cosh(2\eta)}
	\ee^{-4\xi}[1+O(\xi^{-1})],\label{eq:serialcapacityG}\\
	\lambda_{\mathrm{ex}}&=\frac{16}{\pi}D_0q^2\xi\cosh(2\eta)
	\ee^{-4\xi}[1+O(\xi^{-1})],\label{eq:strongrate}\\
	\frac{\Dm}{D_0}&=\frac{4\pi\xi}{\cosh(2\eta)}
	\ee^{-4\xi}[1+O(\xi^{-1})].
	\label{eq:strongmacro}
\end{align}
The first intraregion rates have scale $8D_0q^2\xi$, so the exchange
mode is exponentially separated from them. This is a geometrically
generated reversible-barrier limit of the type underlying classical
escape-rate asymptotics \cite{Kramers1940}; no temperature-dependent
activation energy has been introduced.

Combining the last two equations gives
\begin{equation}
	\frac{a^2\lambda_{\mathrm{ex}}}{\Dm}
	\longrightarrow16\cosh^2(2\eta).
	\label{eq:sharpconstant}
\end{equation}
In the symmetric case $\eta=0$, the two-storage asymptotics give the
first positive cell-periodic rate and approach the constant $16$ in
\eqref{eq:serialperiodic}. Thus that constant cannot be increased over
all positive $a$-periodic profiles. The symmetric profile also has the
smaller period $a/2$; the statement is about functions periodic on
$[0,a]$. An arbitrarily small first harmonic restores primitive period
$a$ and approaches the same constant as $\eta\to0$.

For a physical thin strip the contrast limit must be taken jointly with
the thinness limit. With $\beta=q\bar h$,
$h_{\max}=O(\bar h\sqrt\xi)$ and
$\max|h'|=O(\beta\xi)$ for bounded $\eta$. The conditions
\begin{equation}
	1\ll\xi,\qquad \beta\xi\ll1
	\label{eq:thinstronglimit}
\end{equation}
keep slopes small and transverse equilibration fast relative to the
intraregion rates in the thickness direction. To retain all low bulk
modes in the one-dimensional description, equilibration across the
width also requires $qb\sqrt\xi\ll1$; this condition is unnecessary
within the exactly width-uniform sector. Exponentially small reduced rates are obtained from
positive resistance asymptotics, not by subtracting an uncontrolled
absolute error from a three-dimensional eigenvalue. The serial bound
itself needs no thinness limit as a statement about
\eqref{eq:thinheat}.

At fixed nonzero $\beta$, \eqref{eq:thinstronglimit} permits only a
finite range of $\xi$. The unbounded contrast limit is a statement
about the reduced operator, or a joint limit of increasingly thin
domains, rather than a uniform three-dimensional approximation at
fixed thickness.

\section{Parallel transport and weak exchange}
\label{sec:parallel}

\subsection{Why separate side reservoirs do not isolate one global mode}

A continuous backbone with one weakly connected side reservoir per cell
has two averaged temperatures. Let $C_b,C_s$ be their heat capacities
per cell of length $a$, and let $g$ be the conductance of a negligible-
capacity neck. The leading long-wave balance is
\begin{align}
	C_b\partial_tT_b&=D_0C_b\partial_x^2T_b-g(T_b-T_s),\nonumber\\
	C_s\partial_tT_s&=g(T_b-T_s).
	\label{eq:sidemodel}
\end{align}
Side-cavity diffusion provides a related setting
\cite{Dagdug2007,Dagdug2014}. Here the reduction follows ordinary Fourier
capacity and flux balances; it requires fast local equilibration and
$|K|a\ll1$ for the retained axial wave numbers.

Set $f=C_s/(C_b+C_s)$ and
$\mu=g(C_b^{-1}+C_s^{-1})$. The spatially uniform contrast decays at
$\mu$, whereas the two long-wave branches are
\begin{equation}
	\lambda_\pm^{\mathrm{side}}(K)=\frac12\left[
	D_0K^2+\mu\ \pm
	\sqrt{(D_0K^2+\mu)^2-4D_0K^2g/C_s}\right].
	\label{eq:sidebranches}
\end{equation}
For fixed $g>0$, the hydrodynamic diffusivity is $(1-f)D_0$, but that
single-field limit requires $D_0K^2\ll\mu$ and is nonuniform as
$g\to0$. At fixed nonzero $K$ in the continuum range,
$\lambda_-^{\mathrm{side}}\sim g/C_s$ as $g\to0$.
Consequently, weak attachments leave a family of slow spatial reservoir
patterns. Selecting the uniform contrast by a periodic preparation
does not make it the only slow mode of the entire specimen.

\begin{figure}[!htbp]
\centering
\includegraphics[width=\linewidth]{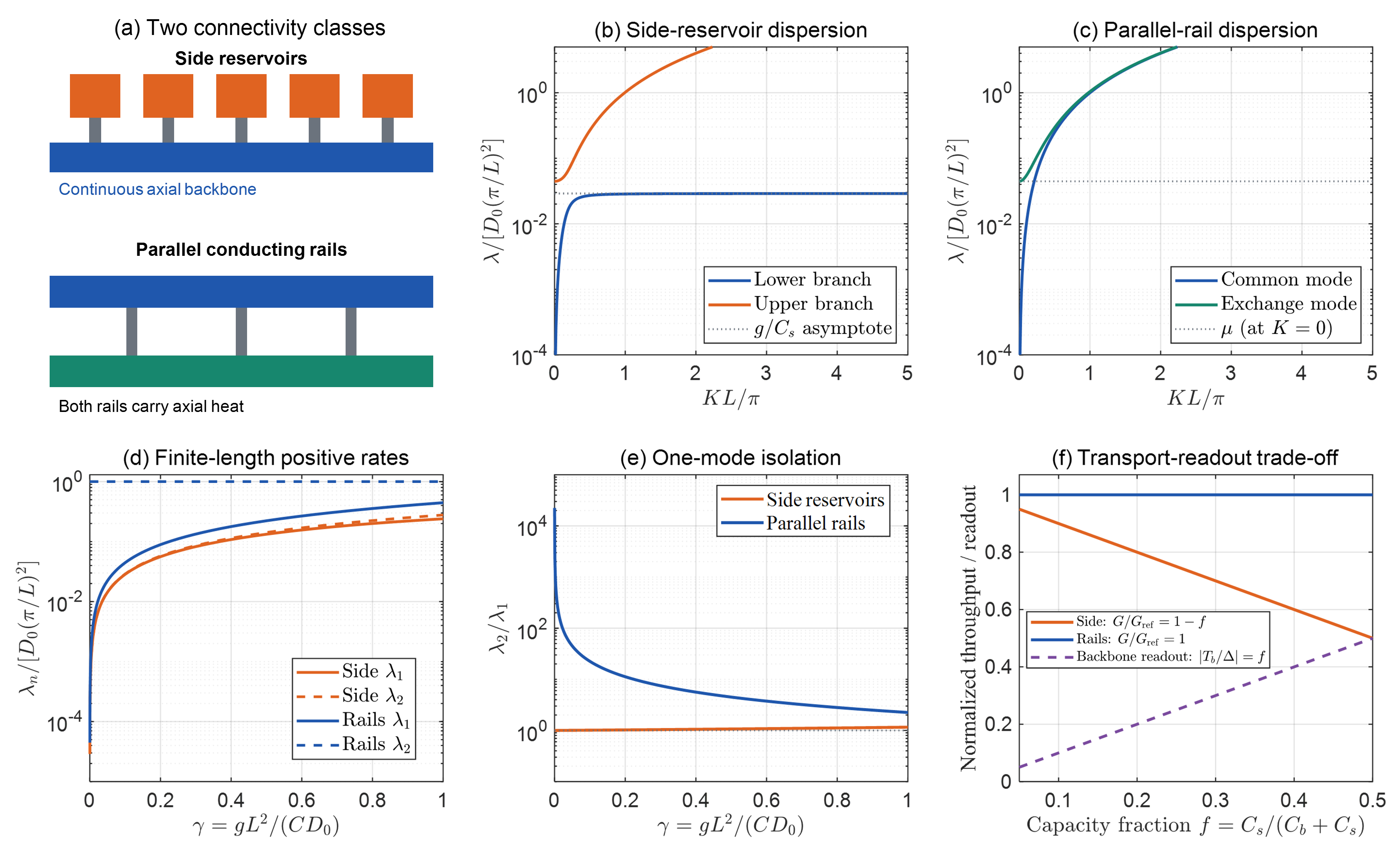}
\caption{\textbf{Connectivity controls the number of slow contrasts.}
(a) A conducting backbone with separate side reservoirs and two
continuous conducting rails. (b) Side-reservoir dispersion, including
the $g/C_s$ slow-branch asymptote. (c) Additive axial and exchange
branches for the two-rail model. (d) First two positive finite-length
rates as the exchange strength varies. (e) Their ratio: separate
reservoirs leave several slow patterns, whereas two continuous rails
isolate one uniform contrast at weak exchange. (f) Leading conductance
fractions and backbone readout of a uniform contrast as the capacity
allocation changes. In these negligible-bridge-volume models,
$\mathcal G/\mathcal G_{\mathrm{ref}}=1-f$ for side reservoirs and
$1$ for conducting rails. The readout shown is relative to $\Delta$.}
\label{fig:fig4}
\end{figure}

To leading order in neglected neck volume, steady transport uses only
the backbone and
\begin{equation}
	\frac{\mathcal G_{\mathrm{ss},D}}{\mathcal G_{\mathrm{ref}}}=1-f.
	\label{eq:sidecost}
\end{equation}
For a zero-mean uniform contrast $\Delta=T_b-T_s$,
$T_b=f\Delta$ and $T_s=-(1-f)\Delta$. Preserving conductance by taking
$f\ll1$ also reduces its backbone readout. A separate storage path that
conducts axially (Fig.~\ref{fig:fig4}) removes both the multiplicity of unequilibrated side
patterns and this direct nonconducting-volume penalty.

\subsection{A capacity-weighted exchange graph}

Consider $r$ continuous axial rails of the same material. Rail $i$ has
heat capacity $C_i$ per cell and exchanges heat with rail $j$ through
conductance $g_{ij}=g_{ji}\geq0$. Define
\begin{equation}
	\mathsf C=\operatorname{diag}(C_1,\ldots,C_r),\qquad
	(\mathsf L_g)_{ii}=\sum_{j\ne i}g_{ij},\qquad
	(\mathsf L_g)_{ij}=-g_{ij}\quad(i\ne j).
	\label{eq:graphdefinitions}
\end{equation}
When bridge storage and access corrections are negligible, the balance is
\begin{equation}
	\mathsf C\partial_t\bm T=D_0\mathsf C\partial_x^2\bm T-\mathsf L_g\bm T.
	\label{eq:railnetwork}
\end{equation}
Its capacity norm dissipates according to
\begin{equation}
	\frac{\dd}{\dd t}\frac1{2a}\int\bm T^\mathsf T\mathsf C\bm T\,\dd x
	=-\frac1a\int\left[
	D_0(\partial_x\bm T)^\mathsf T\mathsf C(\partial_x\bm T)
	+\sum_{i<j}g_{ij}(T_i-T_j)^2\right]\dd x.
	\label{eq:graphdissipation}
\end{equation}
Both channels are passive. The exchange eigenproblem is
\begin{equation}
	\mathsf L_g\bm v_j=\mu_j\mathsf C\bm v_j,
	\qquad 0=\mu_0<\mu_1\leq\cdots\leq\mu_{r-1},
	\label{eq:grapheigen}
\end{equation}
where the exchange graph is assumed connected. The matrix
$\mathsf C^{-1}\mathsf L_g$ is self-adjoint in the capacity inner
product and is the rate operator of a reversible exchange process.

\begin{proposition}[Additive axial and exchange spectra]
	\label{prop:tensorsum}
	For \eqref{eq:railnetwork} on an insulated interval of length $L$,
	\begin{equation}
		\lambda_{m,j}=D_0(m\pi/L)^2+\mu_j,
			\qquad m=0,1,\ldots,\quad j=0,\ldots,r-1.
		\label{eq:tensorspectrum}
	\end{equation}
	If $\mu_1$ is simple and smaller than both $D_0\pi^2/L^2$ and $\mu_2$,
	the first positive rate is $\mu_1$ and the next is
	\begin{equation}
		\lambda_F=\min\{D_0\pi^2/L^2,\mu_2\}.
		\label{eq:graphcompetitor}
	\end{equation}
	For two rails the $\mu_2$ entry is absent.
\end{proposition}

\begin{proof}
	The generator is the tensor sum of the insulated axial operator and
	$\mathsf C^{-1}\mathsf L_g$. Their eigenfunctions are
	$\cos(m\pi x/L)\bm v_j$, giving \eqref{eq:tensorspectrum}. For the
	smallest nonzero exchange rate, all nonuniform axial patterns cost at
	least $D_0\pi^2/L^2$ and all other uniform exchange patterns cost at
	least $\mu_2$. This orders the first two positive rates as stated.
\end{proof}

For two rails, with $C=C_b+C_s$ and $f=C_s/C$, the separated fields are
\begin{equation}
	\overline T=\frac{C_bT_b+C_sT_s}{C},\qquad \Delta=T_b-T_s,
	\qquad \mu=\frac{g}{Cf(1-f)},
	\label{eq:tworailcoordinates}
\end{equation}
and satisfy
\begin{equation}
	\partial_t\overline T=D_0\partial_x^2\overline T,
	\qquad
	\partial_t\Delta=D_0\partial_x^2\Delta-\mu\Delta.
	\label{eq:tworailseparation}
\end{equation}
At fixed finite $L$, $\mu\ll D_0\pi^2/L^2$ isolates one nonconserved
global contrast in this model while retaining $D_0$ for common-
temperature transport. Internal rail and bridge modes omitted from the
model must also decay faster than the target. On an infinite interval,
the common-temperature branch remains gapless as $K\to0$.

\subsection{Independent design of two exchange rates}

For three equal capacities $C_r$, choose conductance $g$ on links
$1$--$2$ and $1$--$3$, and $G$ on link $2$--$3$. The two positive
exchange modes are exactly
\begin{equation}
	\bm v_t\propto(-2,1,1),\quad\mu_t=\frac{3g}{C_r},
	\qquad
	\bm v_f\propto(0,1,-1),\quad\mu_f=\frac{g+2G}{C_r}.
	\label{eq:threeeigen}
\end{equation}
The Jacobian of $(\mu_t,\mu_f)$ with respect to $(g,G)$ has determinant
$6/C_r^2$. Given $\mu_t^*>0$ and $\mu_f^*>\mu_t^*/3$, the inverse rule is
\begin{equation}
	g=\frac{C_r\mu_t^*}{3},\qquad
	G=\frac{C_r}{2}\left(\mu_f^*-\frac{\mu_t^*}{3}\right)>0.
	\label{eq:exchangeinverse}
\end{equation}
Taking $g\ll G$ leaves one weak partition between rail 1 and the other
two, while rapidly relaxing their internal difference. More generally,
the number of small exchange rates reflects the number of weakly
connected groups. Independence here is at allocated rail capacities;
the small bridge-volume corrections require a separate material budget.
If $\mu_t$ is to be the slowest positive exchange rate, the stronger
ordering $\mu_f>\mu_t$, equivalently $G>g$, is required. Global
isolation additionally compares $\mu_t$ with $D_0\pi^2/L^2$.

\subsection{Joint control of longitudinal and exchange spectra}
\label{sec:hybrid}

The same operator framework combines the two-harmonic profile with the
exchange network, rather than treating them as unrelated designs. Let
$r(x)=h(x)/\bar h$, so that $\av r=1$. Assume that every rail has
capacity per unit length $c_i(x)=\bar c_i r(x)$ and axial conductance
coefficient $D_0\bar c_i r(x)$. Also take the transverse conductance
per unit length to be $\gamma_{ij}(x)=r(x)\bar\gamma_{ij}$. The latter
proportionality is an additional design condition, not a consequence of
periodicity. The equations become
\begin{equation}
	\bar c_i r\partial_tT_i
	=D_0\bar c_i\partial_x(r\partial_xT_i)
	-r\sum_j\bar\gamma_{ij}(T_i-T_j).
	\label{eq:hybridbalance}
\end{equation}
After division by $r\bar c_i$, the rate operator is
\begin{equation}
	\mathcal L_{\mathrm{hyb}}=
	\mathcal L_h\otimes\Id+
	\Id\otimes\overline{\mathsf C}^{-1}\mathsf L_{\bar\gamma},
	\qquad \overline{\mathsf C}=\operatorname{diag}(\bar c_i).
	\label{eq:hybridoperator}
\end{equation}
It is a tensor sum in the weighted product space. Consequently, its
Bloch rates are
\begin{equation}
	\lambda_{n,j}(K;\eta,\xi,\bar\gamma)
		=\lambda_n^{\parallel}(K;\eta,\xi)+\mu_j(\bar\gamma).
	\label{eq:hybridspectrum}
\end{equation}
The longitudinal pair splittings are the same in every exchange branch,
and the uniform longitudinal mode has the entire exchange spectrum
$\{\mu_j\}$. Thus longitudinal harmonic controls and transverse
conductance controls act on separate spectral contributions. The
long-wave common branch has \eqref{eq:lifsonjackson}, independently of
the magnitude of a weak transverse exchange rate.

Equation~\eqref{eq:hybridspectrum} makes the relation between local
Bragg-rate design and weak-connectivity isolation precise. A nearly
flat longitudinal profile keeps $\Dm$ close to $D_0$, while a small
$\mu_1$ supplies an arbitrarily slow internal contrast in the reduced
finite specimen. The serial bound remains valid for the longitudinal
operator $\mathcal L_h$, but does not constrain the additional exchange
eigenvalue. The tensor decomposition is exact for
\eqref{eq:hybridbalance}; it is not an exact separation theorem for
general modulated three-dimensional rails. A full bulk realization with
exact separation is constructed in Sec.~\ref{sec:exact}.

\subsection{A fixed-volume bridge construction and bulk variational bounds}
\label{sec:bridges}

To relate the reduced exchange control to material allocation, take two
straight prismatic rails of areas $A_b,A_s$, linked once per cell by a
transverse uniform neck of area $A_n$ and length $\ell_n$. Fix
\begin{equation}
	V_c=a(A_b+A_s)+A_n\ell_n,\qquad
	\varepsilon_n=\frac{A_n\ell_n}{V_c},\qquad
	f=\frac{A_s}{A_b+A_s}.
	\label{eq:bridgebudget}
\end{equation}
As the neck shrinks, its material is returned to the rails, preserving
$V=NV_c$. In the resistance-dominated, negligible-neck-storage limit,
$g=\kappa A_n/\ell_n$ and $C=\cv(1-\varepsilon_n)V_c$, so
\begin{equation}
	\mu=\frac{D_0\varepsilon_n}
	{(1-\varepsilon_n)\ell_n^2 f(1-f)},\qquad
	\frac{D_0\pi^2/L^2}{\mu}
	=\pi^2f(1-f)\left(\frac{\ell_n}{L}\right)^2
	\frac{1-\varepsilon_n}{\varepsilon_n}.
	\label{eq:bridgeexchange}
\end{equation}
The neck diffusion time $\ell_n^2/D_0$ is short relative to the
exchange lifetime when $\varepsilon_n\ll f(1-f)$. The finite-length
factor in \eqref{eq:bridgeexchange} specifies how small the bridges
must be to isolate the uniform contrast in the rail model.

For the full bulk steady problem with complete terminal contacts, the
following bound does not require the reduced exchange approximation:
\begin{equation}
	1-\varepsilon_n\leq
		\frac{\mathcal G_{\mathrm{ss},D}^{\mathrm{bulk}}}{\mathcal G_{\mathrm{ref}}}
		\leq1.
	\label{eq:bridgeconductance}
\end{equation}
Indeed, at a fixed terminal temperature difference, restricting the
Dirichlet energy to the uninterrupted rail subdomains gives at least
$\kappa(A_b+A_s)(T_R-T_L)^2/L$. The affine trial temperature in the
whole domain gives at most $\kappa V(T_R-T_L)^2/L^2$ by the minimum-
dissipation principle. These establish the lower and upper conductance
bounds, respectively.

A bulk insulated trial field that is constant in each rail and linear
across each neck also gives
\begin{equation}
	\Lambda_1^{N,\mathrm{bulk}}\leq
	\frac{D_0\varepsilon_n}
	{(1-\varepsilon_n)\ell_n^2 f(1-f)}.
	\label{eq:bridgebulkupper}
\end{equation}
To see this, give the two rails a temperature difference $\Delta$ and
subtract the mean over the full body. The rail contribution to the
capacity variance is at least
$NCf(1-f)\Delta^2$, irrespective of the neck contribution to the mean;
the trial dissipation is $Ng\Delta^2$. Their quotient proves
\eqref{eq:bridgebulkupper}. This upper bound is valid for the specified
piecewise prismatic domain, even when $g$ is viewed only as its trial
neck conductance rather than an exact reduced exchange coefficient.

Equations~\eqref{eq:bridgeconductance} and \eqref{eq:bridgebulkupper}
prove coexistence of asymptotically preserved bulk conductance and a
vanishing positive bulk rate. They do not prove a positive lower bound
for the second bulk rate of an arbitrary discrete-bridge array. That
remaining issue is settled below for a different, explicitly separable
three-dimensional family.

\section{An exact three-dimensional realization of mode isolation}
\label{sec:exact}

\subsection{A connected transverse dumbbell with fixed total area}

Let $S_1,S_2\subset\mathbb R^2$ be bounded connected regular chambers,
with locally flat facing ports, joined by a straight rectangular neck
of fixed length $\ell$ and width $\epsilon>0$. Take the interior of the
joined closed pieces so that the resulting planar domain
$\widetilde S_\epsilon$ is open and connected. The chambers are fixed
and have no additional narrowing parts. The junctions belong to the
standard planar dumbbell family used in Neumann thin-channel spectral
convergence \cite{Arrieta1995}. The spatial thin-handle theory in
\cite{Gadylshin2005} is related, but the planar result is the one
needed for this cross section.

\begin{figure}[!htbp]
\centering
\includegraphics[width=\linewidth]{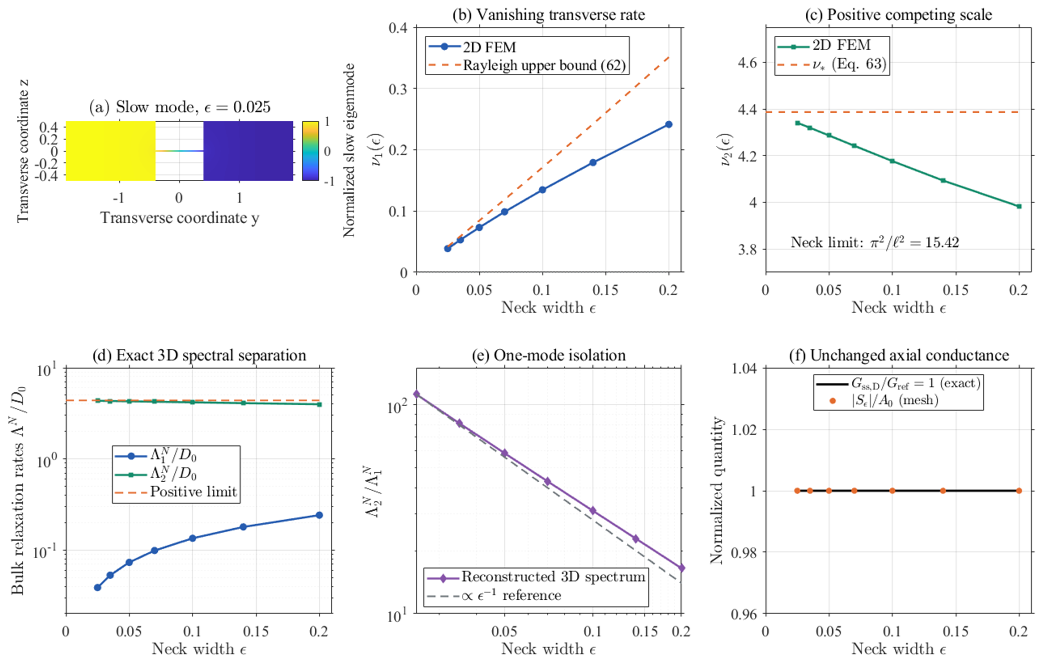}
\caption{\textbf{A separable three-dimensional realization.}
(a) Slow transverse contrast on the connected two-chamber section.
(b) First positive transverse eigenvalue and its Rayleigh upper bound.
(c) Second transverse eigenvalue compared with the positive limiting
scale $\nu_*$. The neck Dirichlet scale is included when taking this
minimum. (d) First two bulk rates reconstructed from the exact product
spectrum \eqref{eq:exactproductspectrum}. (e) Their separation ratio
increases as the neck narrows. (f) Exact axial conductance and sectional
area normalization. The transverse finite-element results shown in the
supplied figure illustrate the construction; the bulk separation and
axial conductance follow analytically from the prismatic geometry.}
\label{fig:fig5}
\end{figure}

Write $A_0=|S_1|+|S_2|$ and normalize the area by a transverse dilation,
\begin{equation}
	d_\epsilon=\left(\frac{A_0}{|\widetilde S_\epsilon|}\right)^{1/2},
	\qquad S_\epsilon=d_\epsilon\widetilde S_\epsilon,
	\qquad |S_\epsilon|=A_0.
	\label{eq:areanormalization}
\end{equation}
Since $d_\epsilon\to1$, this regular rescaling preserves the limiting
chambers and neck length, and multiplies every transverse eigenvalue by
$d_\epsilon^{-2}$. It removes the small neck-area change without a
second geometric limit. Extrude the section over a fixed finite length:
\begin{equation}
	\Omega_\epsilon=(0,L)\times S_\epsilon,\qquad V=LA_0.
	\label{eq:productdomain}
\end{equation}
Figure~\ref{fig:fig5} illustrates the resulting product construction.
Each member with $\epsilon>0$ is a connected three-dimensional conductor
with the same $V,L,\kappa,\cv$.

Let the transverse Neumann eigenvalues be
\begin{equation}
	-\Delta_\perp\phi_j=\nu_j(\epsilon)\phi_j,
	\qquad \partial_{n_\perp}\phi_j=0,
	\qquad 0=\nu_0<\nu_1\leq\nu_2\leq\cdots.
	\label{eq:transversespectrum}
\end{equation}
The insulated bulk spectrum separates exactly:
\begin{equation}
	\Lambda_{m,j}^{N}(\epsilon)
	=D_0\left[(m\pi/L)^2+\nu_j(\epsilon)\right],
	\qquad m=0,1,\ldots.
	\label{eq:exactproductspectrum}
\end{equation}
Unlike the rail model, this formula includes all transverse and
neck-localized modes. It assumes no fast transverse equilibration.

\subsection{One vanishing rate and a positive competing scale}

A chamber-constant trial field, continued linearly through the neck and
then made mean zero, gives
\begin{equation}
	\nu_1(\epsilon)\leq
	\frac{w_\epsilon}{\ell_\epsilon}
	\left(\frac1{A_{1,\epsilon}}+\frac1{A_{2,\epsilon}}\right)
	\longrightarrow0.
	\label{eq:transverseupper}
\end{equation}
Here $w_\epsilon=d_\epsilon\epsilon$,
$\ell_\epsilon=d_\epsilon\ell$ and
$A_{i,\epsilon}=d_\epsilon^2|S_i|$. The proof is the same variance and
neck-energy estimate used in \eqref{eq:bridgebulkupper}. It is an upper
bound, not an asserted exact exchange prefactor.

For these fixed-length rectangular necks one can also obtain the
matching leading lower bound:
\begin{equation}
 \nu_1(\epsilon)=\frac{w_\epsilon}{\ell_\epsilon}
 \left(\frac1{A_{1,\epsilon}}+\frac1{A_{2,\epsilon}}\right)
 [1+o(1)].
 \label{eq:transverseasymptotic}
\end{equation}
The proof in Appendix~\ref{app:neckproof} uses chamber Poincar\'e
inequalities, a shrinking-port trace estimate, and a lower bound for the
neck energy. In two dimensions the port-access correction has a
logarithmic scale, which is subleading to the $\ell_\epsilon/w_\epsilon$
resistance of a neck of fixed positive length. Formula
\eqref{eq:transverseasymptotic} is not asserted for a simultaneous
short-neck limit.

The normalized slow transverse eigenfunction is asymptotically constant
in each chamber, with opposite signs. Taking $A_i=|S_i|$, its chamber
limits in $L^2$ are, up to an overall sign,
\begin{equation}
 \phi_1|_{S_1}\ \longrightarrow\sqrt{\frac{A_2}{A_1(A_1+A_2)}},\qquad
 \phi_1|_{S_2}\ \longrightarrow-\sqrt{\frac{A_1}{A_2(A_1+A_2)}}.
 \label{eq:chambercontrastshape}
\end{equation}
The dilation is understood when comparing these functions on fixed
chambers. For the extruded body, define chamber capacities and a leading
exchange conductance by
\begin{equation}
 C_i=\cv L A_{i,\epsilon},\qquad
 g_\perp=\frac{\kappa Lw_\epsilon}{\ell_\epsilon}.
 \quad\text{Then}\quad
 \Lambda_1^N=g_\perp(C_1^{-1}+C_2^{-1})[1+o(1)].
 \label{eq:exactgraphrate}
\end{equation}
This supplies the capacity-weighted two-rail exchange law directly from
the bulk operator. Its lifetime grows as $w_\epsilon^{-1}$ at fixed
chamber areas and neck length.

For this regular constant-width thin-channel family, the limiting
transverse spectrum consists, with multiplicities, of the two chamber
Neumann spectra and the Dirichlet interval spectrum of the collapsed
neck \cite{Arrieta1995,JimboKosugi2009}. The two chamber constants produce one exact
zero and one vanishing eigenvalue. The next eigenvalue therefore obeys
\begin{equation}
	\nu_2(\epsilon)\longrightarrow\nu_*:=
	\min\left\{\nu_1^N(S_1),\nu_1^N(S_2),\frac{\pi^2}{\ell^2}\right\}>0.
	\label{eq:transverselimit}
\end{equation}
The notation $\nu_1^N(S_i)$ denotes the first positive eigenvalue of
$-\Delta_\perp$ in chamber $i$, with dimensions of inverse length
squared. The neck contribution must be retained: a long narrow
connector can support a lower competing scale than either chamber.
For a nonuniform collapsed neck, its appropriate one-dimensional
Dirichlet spectrum replaces $\pi^2/\ell^2$.
The positive lower limit can also be proved without identifying the
entire limiting spectrum: any normalized sequence with vanishing
Dirichlet energy has vanishing neck mass and reduces to two chamber
constants. There is room for only two orthogonal such states, one of
which is the conserved constant. Appendix~\ref{app:neckproof} gives
this compactness argument. The explicit value $\nu_*$ uses the
stronger thin-channel spectral convergence result.

\begin{proposition}[Bulk isolation at unchanged axial conductance]
	\label{prop:bulk}
	For the fixed-volume prismatic family \eqref{eq:productdomain},
	\begin{equation}
		\frac{\mathcal G_{\mathrm{ss},D}(\epsilon)}{\mathcal G_{\mathrm{ref}}}=1,
			\qquad \Lambda_1^N(\epsilon)\to0,
			\qquad \Lambda_2^N(\epsilon)\to D_0
			\min\{\pi^2/L^2,\nu_*\}>0.
		\label{eq:triplecompatibility}
	\end{equation}
	The positive bulk rates are ordered with multiplicity. The limiting
	statement holds at fixed $L,\ell$ and nondegenerate chambers.
\end{proposition}

\begin{proof}
	For sufficiently small $\epsilon$, \eqref{eq:exactproductspectrum}
	orders the first two positive rates as
	\begin{equation}
		\Lambda_1^N=D_0\nu_1,\qquad
		\Lambda_2^N=D_0\min\{\pi^2/L^2,\nu_2\}.
		\label{eq:orderedbulk}
	\end{equation}
	The limits follow from \eqref{eq:transverseupper} and
	\eqref{eq:transverselimit}. For the distinct terminally contacted
	steady problem, the full solution is
	\begin{equation}
		T_{\mathrm{ss}}(x,\bm y)=T_L+\frac{T_R-T_L}{L}x,
		\qquad
		\mathcal G_{\mathrm{ss},D}=\frac{\kappa A_0}{L}
		=\frac{\kappa V}{L^2}.
		\label{eq:affineconductance}
	\end{equation}
	The lateral normal has no axial component, so this solution satisfies
	all lateral insulation conditions. Uniqueness of the steady mixed
	boundary-value problem completes the conductance statement.
\end{proof}

The relevant factorization is exact in the full bulk operator:
axial transport acts on one coordinate while the slow exchange acts on
the transverse cross section. The conductance statement uses the
complete connected terminal section, including both chambers.
Contacting only one chamber would impose a different mixed boundary
problem and need not preserve the reference conductance. The result
compares two protocols on the same geometry: fully contacted axial
throughput and fully insulated retention after preparation.
In fact, the sectional mean
\begin{equation}
	\overline T(x,t)=\frac1{A_0}\int_{S_\epsilon}T(x,\bm y,t)\,\dd A
	\quad\text{satisfies}\quad
	\partial_t\overline T=D_0\partial_x^2\overline T
	\label{eq:exactsectionmean}
\end{equation}
without a transverse-equilibration assumption. Thus the common axial
diffusivity is also exactly $D_0$, although differential transverse
equilibration becomes arbitrarily slow.

The proof uses established thin-channel convergence for
\eqref{eq:transverselimit}. Its use here is the fixed-volume,
fixed-length compatibility statement with exact axial conduction; no
new general thin-neck convergence theorem is assumed. Additional weak
chamber partitions would introduce additional small eigenvalues and
invalidate one-mode isolation. The limiting disconnected domain is a
mathematical limit; every operating member remains connected.

\subsection{Metastable reduction and noncommuting long-time limits}

The isolated bulk mode has a direct two-compartment interpretation.
To leading order in the neck width, the chamber-average deviations obey
\begin{equation}
 C_1\dot\Theta_1=-g_\perp(\Theta_1-\Theta_2),\qquad
 C_2\dot\Theta_2=g_\perp(\Theta_1-\Theta_2).
 \label{eq:bulkcompartments}
\end{equation}
Writing $C_\Sigma=C_1+C_2$ and $\Delta=\Theta_1-\Theta_2$, the
mean-zero constraint gives $\Theta_1=C_2\Delta/C_\Sigma$ and
$\Theta_2=-C_1\Delta/C_\Sigma$. The corresponding quadratic
nonuniformity and its dissipation are
\begin{equation}
 \mathcal V=\frac12\frac{C_1C_2}{C_\Sigma}\Delta^2,\qquad
 \dot{\mathcal V}=-g_\perp\Delta^2=-2\Lambda_t\mathcal V,\qquad
 \Lambda_t=g_\perp(C_1^{-1}+C_2^{-1}).
 \label{eq:contrastdissipation}
\end{equation}
These equalities are exact for the two-compartment reduction;
$\Lambda_t=\Lambda_1^N[1+o(1)]$ for the bulk family. The temperature
contrast decays at $\Lambda_t$, whereas its quadratic entropy deficit
decays at $2\Lambda_t$. A long lifetime therefore reflects suppressed
entropy production for this spatial pattern, not suppressed axial
conduction for a common temperature gradient.

For a normalized initial slow eigenmode, define its normalized modal
amplitude $A_\epsilon(t)=\exp[-\Lambda_1^N(\epsilon)t]$. Then
\begin{equation}
 \lim_{\epsilon\to0}\lim_{t\to\infty}A_\epsilon(t)=0,
 \qquad
 \lim_{t\to\infty}\lim_{\epsilon\to0}A_\epsilon(t)=1.
 \label{eq:noncommutinglimits}
\end{equation}
Every connected member equilibrates, but taking the disconnected limit
first conserves the two chamber heats separately. At small positive
width the physically relevant intermediate regime is
$(\Lambda_2^N)^{-1}\ll t\ll(\Lambda_1^N)^{-1}$, provided the
preparation excites the contrast. This is a metastable interval, rather
than a new thermodynamic equilibrium phase.

\subsection{Boundary conditions and the order of limits}

If the terminals remain clamped after preparation, perturbations around
the steady state have homogeneous axial Dirichlet conditions. The bulk
rates then are
\begin{equation}
	\Lambda_{m,j}^{D}=D_0[(m\pi/L)^2+\nu_j],\qquad m=1,2,\ldots,
	\label{eq:dirichletfloor}
\end{equation}
and have the floor $D_0\pi^2/L^2$. The vanishing insulated contrast rate
in \eqref{eq:triplecompatibility} cannot be assigned to that continuously
contacted experiment.

Similarly, the positive competing limit is not uniform in specimen
length. From \eqref{eq:orderedbulk},
\begin{equation}
	\mathcal S_L(\epsilon):=\frac{\Lambda_2^N}{\Lambda_1^N}
	=\frac{\min\{\pi^2/L^2,\nu_2(\epsilon)\}}{\nu_1(\epsilon)}.
	\label{eq:exactselectivity}
\end{equation}
At fixed $L$, $\mathcal S_L\to\infty$. In a simultaneous limit
$L\to\infty$, global isolation instead requires
$L^2\nu_1(\epsilon)\to0$, as well as separation from the transverse
competing scales. On an infinite axial domain there are arbitrarily
slow common-temperature long waves. The result concerns a finite
insulated specimen or a specified spatial sector.

\section{Observable relaxation and the selection window}
\label{sec:observation}

\subsection{A spectral projection criterion for a retained contrast}

\begin{figure}[!htbp]
\centering
\includegraphics[width=\linewidth]{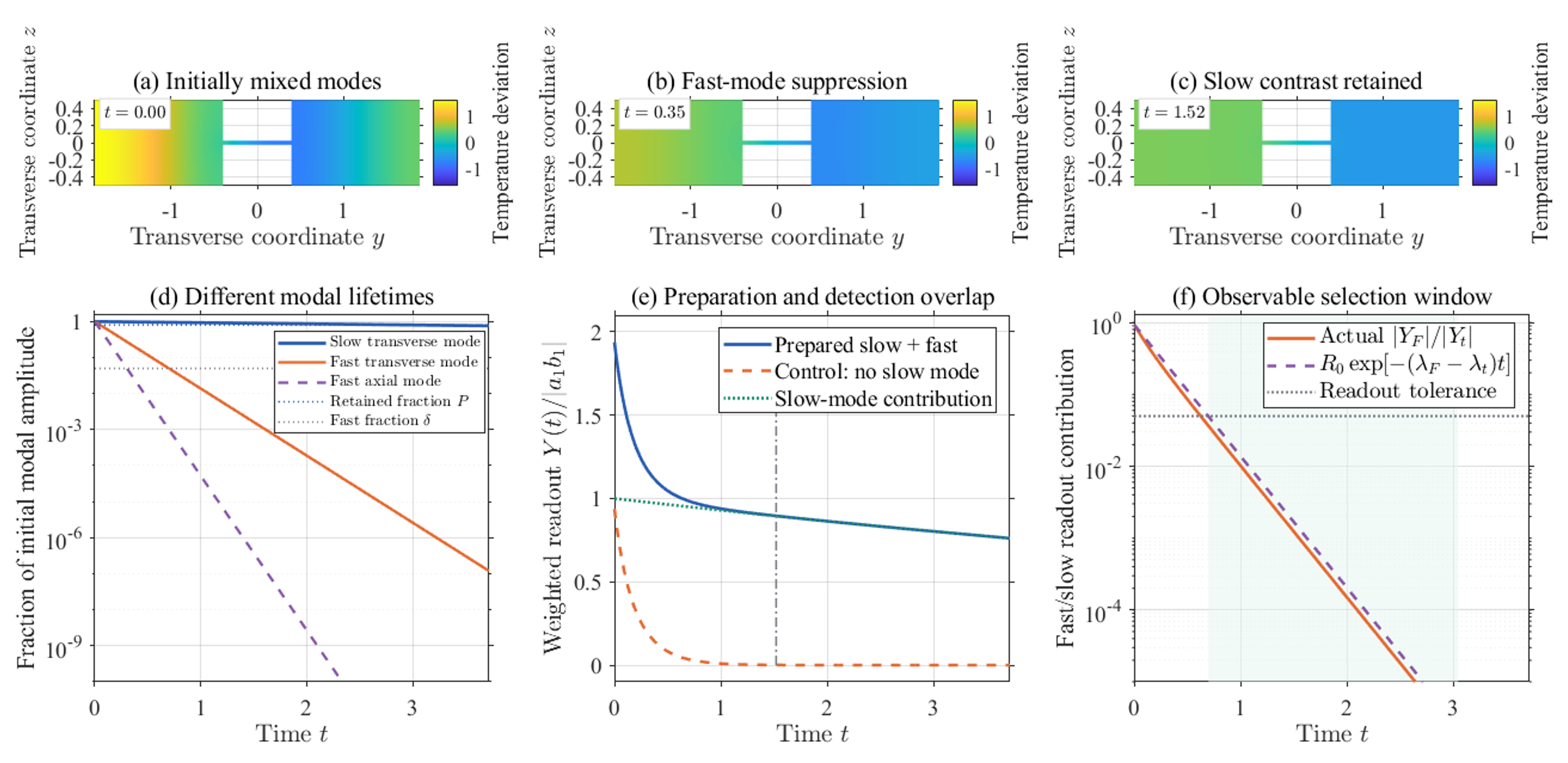}
\caption{\textbf{Preparing and observing a retained thermal contrast.}
(a--c) Temperature fields evolving from a mixed preparation through
fast-mode suppression to a retained chamber contrast. (d) Decay of slow
transverse, fast transverse and fast axial amplitudes, with retention
and suppression thresholds. (e) Weighted readout with and without a
slow-mode preparation component, compared with the target contribution.
(f) Fast-to-target readout magnitude and its overlap-weighted spectral
bound. The shaded interval illustrates simultaneous fast-signal
suppression and target retention. Equations~\eqref{eq:puritybound}
and \eqref{eq:overlapwindow} distinguish modal separation from the
purity of a particular measured signal.}
\label{fig:fig6}
\end{figure}

Let $u_n$ be orthonormal in the relevant capacity inner product. For
the full bulk field, the strip or the rail model, a mean-zero
preparation has the expansion
\begin{equation}
	\theta(t)=\sum_{n\geq1}a_nu_n\ee^{-\lambda_nt},
	\qquad a_n=(u_n,\theta(0))_C.
	\label{eq:modalexpansion}
\end{equation}
If the first positive eigenvalue is simple and $P_1$ projects onto its
eigenfunction, self-adjointness gives the exact error bound
\begin{equation}
	\left\|\theta(t)-a_1u_1\ee^{-\lambda_1t}\right\|_C
	\leq\ee^{-\lambda_2t}
	\left\|(\Id-P_1)\theta(0)\right\|_C.
	\label{eq:projectionbound}
\end{equation}
This is the dynamical consequence of strong spectral isolation
(Fig.~\ref{fig:fig6}). After
the fast components have decayed, the remaining nonuniform field is
approximately one capacity-weighted contrast.

For a linear temperature readout $Y(t)=(w,\theta(t))_C$,
\begin{equation}
	Y(t)=\sum_{n\geq1}a_nb_n\ee^{-\lambda_nt},\qquad b_n=(w,u_n)_C.
	\label{eq:readout}
\end{equation}
The target is observable only if both preparation and readout overlaps
are nonzero. A capacity-weighted mean readout has zero overlap with
every positive-rate mode. Differential rail heating and a rail-resolved
temperature difference access the exchange mode directly, whereas
uniform heating of both rails does not excite it.

Let $P\in(0,1)$ be a required retained fraction of the target amplitude,
and let $\delta\in(0,1)$ bound the remaining fraction of each competing
amplitude relative to its own initial value. With
$\lambda_F$ the smallest relevant competing rate, a sufficient window is
\begin{equation}
	\frac{\ln(1/\delta)}{\lambda_F}\leq t_{\mathrm{obs}}
	\leq\frac{\ln(1/P)}{\lambda_t},\qquad
	\frac{\lambda_F}{\lambda_t}\geq
	\frac{\ln(1/\delta)}{\ln(1/P)}.
	\label{eq:amplitudewindow}
\end{equation}
This condition alone does not guarantee output purity for arbitrary
initial amplitude ratios. A readout-specific bound is obtained by
setting $R_0=\sum_{j\in\mathcal F}|a_jb_j|/|a_tb_t|$, assumed finite.
If $\lambda_F>\lambda_t$, then
\begin{equation}
	\frac{|Y_{\mathcal F}(t)|}{|a_tb_t|\ee^{-\lambda_tt}}
	\leq R_0\ee^{-(\lambda_F-\lambda_t)t}.
	\label{eq:puritybound}
\end{equation}
The spectrum specifies the available temporal separation; the overlaps
determine whether the selected contrast is actually prepared and seen.

A norm bound avoids the need to resolve each fast-mode overlap.
For bounded linear readout, $w\in L^2_C$, let $P_{\mathcal F}$ project
onto all non-target mean-zero modes, whose rates are at least
$\lambda_F>\lambda_t$. Cauchy--Schwarz and the spectral theorem give
\begin{equation}
 R_0\leq B_0:=
 \frac{\|P_{\mathcal F}\theta(0)\|_C
       \|P_{\mathcal F}w\|_C}{|a_tb_t|},\qquad
 \frac{|Y_{\mathcal F}(t)|}{|a_tb_t|\ee^{-\lambda_tt}}
 \leq B_0\ee^{-(\lambda_F-\lambda_t)t}.
 \label{eq:normpurity}
\end{equation}
Here $a_tb_t\ne0$ is essential. Ideal point evaluation is not a bounded
functional on $L^2_C$; the criterion applies directly to finite-volume
readout weights, or at positive time after appropriate regularity
estimates. A surface sensor requires a trace estimate or a specified
depth average before this norm bound is used.
Given an output-contamination tolerance $\varepsilon_Y\in(0,1)$ and
a retained target fraction $P$, a sufficient observation interval is
\begin{equation}
 \frac{[\ln(B_0/\varepsilon_Y)]_+}{\lambda_F-\lambda_t}
 \leq t_{\mathrm{obs}}\leq\frac{\ln(1/P)}{\lambda_t},\qquad
 \frac{\lambda_F}{\lambda_t}\geq
 1+\frac{[\ln(B_0/\varepsilon_Y)]_+}{\ln(1/P)}.
 \label{eq:overlapwindow}
\end{equation}
We use $[z]_+=\max\{z,0\}$, and take the lower endpoint to be zero
when $B_0=0$. This bound controls the magnitude of the total fast
contribution relative to the target signal; it is stronger than
requiring every fast amplitude merely to decay relative to its own
initial value. It is sufficient rather than necessary, since fast-mode
cancellations can improve an actual readout.

\subsection{Reversible correlations and retained contrast}

The same spectral statements have a statistical-mechanical formulation.
Use the invariant probability measure $\pi$ of the corresponding
reversible diffusion: uniform volume measure for the homogeneous bulk,
$h\,\dd x/M$ for the strip, or normalized capacities for the exchange
graph. For a mean-zero observable $F$ and an orthonormal basis
$\varphi_n$ in $L^2(\pi)$, its stationary diffusion correlation is
\begin{equation}
	C_F(t)=\av{F,\ee^{-t\mathcal L}F}_\pi
	=\sum_{n\geq1}|\av{\varphi_n,F}_\pi|^2\ee^{-\lambda_nt}.
	\label{eq:reversiblecorrelation}
\end{equation}
This is an exact identity for the auxiliary reversible diffusion
process generated by the thermal rate operator. It does not posit a
microscopic stochastic model for temperature fluctuations. A thermal
preparation proportional to $F$, read with the same weight, has the
same normalized relaxation kernel. With nonzero slow-mode overlap,
the isolated contrast supplies an asymptotically single-exponential
long correlation time.
Its normalized integrated correlation time is
\begin{equation}
 \tau_{\mathrm{int}}(F):=\frac1{C_F(0)}\int_0^\infty C_F(t)\,\dd t
 =\sum_{n\geq1}\frac{p_n(F)}{\lambda_n},\qquad
 p_n(F)=\frac{|\av{\varphi_n,F}_\pi|^2}{\|F\|_\pi^2}.
 \label{eq:integratedcorrelation}
\end{equation}
For a family with $\lambda_1\to0$ and $\liminf\lambda_2>0$,
$\tau_{\mathrm{int}}=p_1(F)/\lambda_1+O(1)$.
A merely nonzero overlap at each geometry need not give a divergent
measured correlation time: the ratio $p_1(F)/\lambda_1$ must diverge.
An overlap bounded below by a positive constant is a simple sufficient
condition. Spatially averaged measurements that annihilate the chamber
contrast miss this long time even when the bulk spectrum is isolated.

For two rails the quadratic contrast measure per cell is
\begin{equation}
	\mathcal V_{\mathrm{contrast}}
	=\frac C2 f(1-f)\Delta^2,\qquad
	T_b=f\Delta,\qquad T_s=-(1-f)\Delta.
	\label{eq:contraststorage}
\end{equation}
In the side-reservoir geometry, a conductance-loss allowance
$\varepsilon$ enforces $f\leq\varepsilon$ at leading order, and hence
\begin{equation}
	\frac{|T_b|}{|T_s|}=\frac f{1-f}
	\leq\frac{\varepsilon}{1-\varepsilon}.
	\label{eq:signalbudget}
\end{equation}
In the two-rail geometry both allocations conduct axially, so $f=1/2$
gives equal and opposite signals while retaining
\eqref{eq:bridgeconductance}. The exact prismatic realization similarly
permits comparable chamber volumes without an axial conductance loss.

These formulas identify retention as persistence of a prepared
temperature contrast. Conserved mean heat is present in every insulated
sample and has been excluded from the spectral comparison. A weak
transverse connection creates a long but finite decay time in each
connected body; neither bistable equilibrium states nor nondissipative
information storage follow from the linear Fourier model. Surface heat
losses, finite neck dimensions and imperfect preparation would add
physical decay or readout scales beyond the ideal insulated problems
treated here.

\subsection{Preparation and a finite heat-loss scale}

For a volumetric heating rate $q_{\mathrm{src}}(\bm x,t)$, the forced
bulk equation is $\partial_t\theta=-\mathcal L\theta+
q_{\mathrm{src}}/\cv$. Its positive-mode amplitudes satisfy
\begin{equation}
 \dot a_n+\lambda_na_n=f_n(t),\qquad
 f_n(t)=\int_\Omega u_n^*q_{\mathrm{src}}\,\dd V,\qquad
 a_n(t_p)=\int_0^{t_p}\ee^{-\lambda_n(t_p-s)}f_n(s)\,\dd s,
 \label{eq:forcedpreparation}
\end{equation}
for an initially equilibrated body and preparation time $t_p$.
The conserved mean is subtracted after the pulse. Spatially uniform
heating has $f_1=0$, whereas selective heating of one chamber has a
nonzero contrast projection; active cooling is not required to obtain
opposite deviations from the final mean. For constant $f_1$,
$a_1(t_p)=f_1(1-\ee^{-\lambda_1t_p})/\lambda_1$, tending to
$f_1t_p$ for $\lambda_1t_p\ll1$. Thus weak transverse exchange does
not itself prevent rapid writing by a spatially selective source.

Finite heat leakage supplies another time scale. In a lumped chamber
approximation, let $H_i$ be the conductance from chamber $i$ to a common
ambient bath and set $\gamma_i=H_i/C_i$. The rate matrix becomes
\begin{equation}
 \mathsf M=\begin{pmatrix}
 \gamma_1+g_\perp/C_1&-g_\perp/C_1\\
 -g_\perp/C_2&\gamma_2+g_\perp/C_2
 \end{pmatrix}.
 \label{eq:leakmatrix}
\end{equation}
For matched leakage rates $\gamma_1=\gamma_2=\gamma$, the common
and differential modes decay at $\gamma$ and $\gamma+\Lambda_t$,
respectively. Hence
\begin{equation}
 \tau_{\mathrm{contrast}}=\frac1{\gamma+\Lambda_t},\qquad
 \gamma\ll\Lambda_t\ll\Lambda_F
 \label{eq:losscriterion}
\end{equation}
is the scale ordering needed for geometry-controlled retention together
with fast internal equilibration. At fixed $\gamma>0$, narrowing the
neck eventually saturates the lifetime at $\gamma^{-1}$. Unequal
$\gamma_i$ mix the common and differential coordinates. This model
assumes nearly isothermal chambers and neglects distributed neck loss;
it illustrates a practical limitation, rather than replacing the
insulated bulk theorem by a claim for arbitrary Robin boundaries.

\section{Physical interpretation and scope}
\label{sec:scope}

The results distinguish a material diffusivity, a macroscopic transport
coefficient, and an internal contrast lifetime. The intrinsic $D_0$ is
unchanged throughout. In a serial strip, geometry redistributes both
storage and axial resistance; the same bottlenecks responsible for a
very slow first positive rate also suppress $\Dm$. The bound in
Proposition~\ref{prop:serial} expresses this coupling without a
weak-modulation assumption. Local harmonic design remains useful
because selected rates change at first order while transport changes
only at second order, and an exact local transport constraint can also
be imposed.

The parallel construction changes this allocation of resistance. Each
chamber remains a complete axial conducting path, while a narrow
transverse connection limits differential equilibration. A weak link
then splits the two chamber constants into a conserved common
component and one slowly decaying contrast. Fixed chamber and neck
length scales keep every other transverse mode fast. The exact product
geometry proves the compatibility with unchanged axial conductance,
while the graph and hybrid models describe broader designs only under
their stated reduction and proportionality assumptions. The underlying
thin-neck spectral facts are established results; their role here is
to make this transport--retention comparison precise at fixed material,
volume and specimen length.

Finally, a spectral lifetime becomes a useful retention time only when
preparation and measurement have finite overlap with the contrast.
The bound \eqref{eq:overlapwindow} supplies this operational criterion.
The comparison uses insulated relaxation after preparation and a
separate terminally contacted throughput protocol; maintaining those
contacts during retention changes the decay spectrum. These distinctions
identify which geometrical properties control a selected thermal
response without changing Fourier's constitutive law or implying
stable binary equilibria.

\section{Conclusions}
\label{sec:conclusions}

We have characterized how geometry controls the relaxation spectrum of
a homogeneous Fourier conductor while distinguishing internal
 equilibration from steady heat throughput. In the reduced serial
problem, two logarithmic shape harmonics independently tune selected
Bloch-rate pairs, and a one-parameter redistribution remains possible
on an exact local transport level set. This local freedom does not
remove the global restriction: the weighted Poincar\'e bound requires
vanishing effective diffusivity whenever the first positive insulated
rate vanishes at fixed length. The cell-periodic constant is sharp.

Continuous parallel paths remove this serial restriction by assigning
axial transport and differential exchange to different spatial
sectors. Capacity-weighted rail models identify the mechanism and the
number of slow contrasts. An extruded thin-necked section then gives a
full three-dimensional realization at fixed material properties, volume
and axial length. Its conductance is exactly $\kappa V/L^2$, while
one positive insulated rate tends to zero and the next remains bounded
away from zero. The leading slow rate is set by transverse neck
conductance and inverse chamber capacities; the competing scale must
include axial, chamber and neck modes. The bulk construction uses
established thin-channel spectral convergence rather than asserting a
new general convergence theorem.

Spectral isolation produces a useful thermal response only when both
preparation and readout overlap with the slow contrast. The derived
observation bounds make this dependence explicit, and the forced-mode
balance identifies spatially selective heating as a way to prepare the
contrast. The retained quantity is a nonuniform temperature pattern
with a long but finite lifetime. Its persistence requires insulation
during retention; clamped terminals or finite environmental leakage
introduce additional decay scales. The resulting design principle is
to preserve continuous throughput paths while independently limiting
the exchange that dissipates the chosen internal contrast.

\appendix
\section{Second-order elimination in the two Bloch pairs}
\label{app:perturbation}

Write $H=H_0+V^{(1)}+V^{(2)}$, where $H_0=-\partial_s^2$,
$V^{(1)}=W''$ and $V^{(2)}=(W')^2$. For a free degenerate pair with
eigenvalue $E_*$ and projection $P$, let $Q=\Id-P$. To second order,
\begin{equation}
	H_{\mathrm{eff}}=E_*P+PV^{(1)}P+PV^{(2)}P
	+PV^{(1)}Q(E_*-QH_0Q)^{-1}QV^{(1)}P+O(\rho^3).
	\label{eq:schur}
\end{equation}
The denominator excludes only the resonant pair; all remote momenta
linked at this order must be retained.

For $E_*=1/4$ and momenta $\pm1/2$, the direct second-order
off-diagonal term is $\eta\xi\ee^{\ii\phi}$. The intermediate states
$3/2$ and $-3/2$ each contribute
$-\eta\xi\ee^{\ii\phi}/2$ because their denominators are
$1/4-9/4=-2$. The net second-order off-diagonal entry is zero. The
common diagonal correction is
\begin{equation}
	\frac{\eta^2}{2}+2\xi^2-\frac{\eta^2}{8}
	-\frac{4\xi^2}{2}-\frac{4\xi^2}{6}
	=\frac{3\eta^2}{8}-\frac{2\xi^2}{3}.
	\label{eq:appendixdiag1}
\end{equation}
For $E_*=1$ and momenta $\pm1$, the direct term $-\eta^2/4$ is
canceled by the path $+1\to0\to-1$, whose denominator is $1-0=1$.
The common diagonal correction becomes
\begin{equation}
	\frac{\eta^2}{2}+2\xi^2+\frac{\eta^2}{4}
	-\frac{\eta^2}{12}-\frac{4\xi^2}{8}
	=\frac{2\eta^2}{3}+\frac{3\xi^2}{2}.
	\label{eq:appendixdiag2}
\end{equation}
These cancellations rely on $U=W''+(W')^2$ and need not hold for an
arbitrarily prescribed scalar potential. They establish
\eqref{eq:effectivepair1} and \eqref{eq:effectivepair2} for the leading
thin-layer Fourier operator.

\section{Storage and resistance in the strong serial limit}
\label{app:asymptotics}

For $W=\eta\cos s+\xi\cos2s$, the storage maxima at $0,\pi$ give
\begin{align}
	Z&=\frac{\ee^{2\xi}}{2\pi}
	\left[\ee^{2\eta}\sqrt{\frac{\pi}{4\xi+\eta}}
	+\ee^{-2\eta}\sqrt{\frac{\pi}{4\xi-\eta}}\right]
	[1+O(\xi^{-1})]\nonumber\\
	&=\frac{\ee^{2\xi}\cosh(2\eta)}{2\sqrt{\pi\xi}}
	[1+O(\xi^{-1})].
	\label{eq:Zasymptotic}
\end{align}
Dividing the two peak integrals by $Z$ yields the capacities in
\eqref{eq:serialcapacityG}. The passages have
\begin{equation}
	\cos s_n=-\frac{\eta}{4\xi},\qquad
	W(s_n)=-\xi-\frac{\eta^2}{8\xi},\qquad
	W''(s_n)=4\xi-\frac{\eta^2}{4\xi}.
	\label{eq:neckstationary}
\end{equation}
The resistance of either path between adjacent peaks is therefore
\begin{equation}
	\int_{A\to B}\frac{\dd x}{h(x)}
	=\frac Z{q\bar h}\int_{A\to B}\ee^{-2W(s)}\,\dd s
	=\frac{\cosh(2\eta)}{4q\bar h\xi}\ee^{4\xi}
	[1+O(\xi^{-1})],
	\label{eq:pathresistance}
\end{equation}
which supplies $G$. The temperature generator around either peak is
\begin{equation}
	-\mathcal L_h\simeq D_0q^2
	\left[\partial_\delta^2-2(4\xi\pm\eta)\delta\partial_\delta\right],
	\label{eq:OUgenerator}
\end{equation}
and its first local nonzero rate has scale $8D_0q^2\xi$.
Storage is concentrated near these peaks, while resistance is
concentrated in the separating passages. Eliminating the fast
intraregion modes gives the two-capacity exchange balance
\eqref{eq:serialreservoirs}; its slow Rayleigh quotient is
$2G(C_A^{-1}+C_B^{-1})$ to leading order. This establishes
\eqref{eq:strongrate} within the reduced high-contrast limit. Finally,
\begin{equation}
	\av{\ee^{-2W}}_s=
	\frac{\ee^{2\xi}}{2\sqrt{\pi\xi}}[1+O(\xi^{-1})]
	\label{eq:inverseaverage}
\end{equation}
combined with \eqref{eq:Zasymptotic} proves
\eqref{eq:strongmacro} directly from positive integrals.

Sharpness of the periodic bound can also be verified variationally,
without relying on the exchange reduction. At $\eta=0$, choose a trial
field equal to $+1$ and $-1$ in shrinking neighbourhoods of the two
storage maxima and linear in the resistance coordinate on each
connecting path. Choose these neighbourhoods wider than
$\xi^{-1/2}$ but narrower than a cell, so they contain asymptotically
all storage and asymptotically no resistance. Symmetry ensures zero
weighted mean. Each path has resistance $R/2[1+o(1)]$ and temperature
drop $2$, so
\begin{equation}
	\int h(u')^2\,\dd x=\frac{16}{R}[1+o(1)],\qquad
	\int hu^2\,\dd x=M[1+o(1)].
	\label{eq:sharpnesstrial}
\end{equation}
The Rayleigh upper bound and Proposition~\ref{prop:serial} then imply
$a^2\lambda_1^{\mathrm{per}}/\Dm\to16$.

\section{Thin-neck exchange rate and exclusion of a second slow contrast}
\label{app:neckproof}

We give the leading-rate and compactness arguments for the planar family
in Sec.~\ref{sec:exact}. It is enough to work on the undilated section
$\widetilde S_\epsilon$, with chambers of areas $A_1,A_2$ and the
rectangular neck $R_\epsilon=(0,\ell)\times(0,\epsilon)$.
The area-normalizing dilation subsequently multiplies rates by
$d_\epsilon^{-2}$. Constants below may depend on the fixed chambers,
ports and $\ell$, but not on $\epsilon$.

For $u\in H^1(\widetilde S_\epsilon)$, let $U_i$ denote its average in
chamber $i$, $p_i$ its average on the corresponding port, and
$E_i=\int_{S_i}|\nabla u|^2\,\dd A$. Fixed-chamber Poincar\'e
inequalities give
\begin{equation}
 \int_{S_i}|u-U_i|^2\,\dd A\leq K_i E_i.
 \label{eq:chamberpoincare}
\end{equation}
At a locally flat planar port of width $\epsilon$, the average trace
satisfies
\begin{equation}
 |p_i-U_i|^2\leq
 K\left[1+\ln(r_*/\epsilon)\right]E_i,
 \qquad 0<\epsilon<r_*.
 \label{eq:porttrace}
\end{equation}
Here $r_*>0$ is a fixed local length. One obtains the logarithm by
comparing averages over nested half-discs from radius $\epsilon$ to
$r_*$: Cauchy--Schwarz applied to radial derivatives produces
$\int_\epsilon^{r_*}\dd r/r$. A trace estimate on the smallest
rescaled half-disc compares its average with the port average; the
fixed outer region is controlled by \eqref{eq:chamberpoincare}.
This explains both the dimensionless logarithm and its uniformity for
the fixed locally flat junctions considered here.

Let $v(z)=\epsilon^{-1}\int_0^\epsilon u(z,y)\,\dd y$ be the neck
average. Jensen's inequality and one-dimensional Cauchy--Schwarz imply
\begin{equation}
 \int_{R_\epsilon}|\nabla u|^2\,\dd A
 \geq\epsilon\int_0^\ell|v'|^2\,\dd z
 \geq\frac\epsilon\ell|p_1-p_2|^2.
 \label{eq:necklower}
\end{equation}
Decomposing $u=v+(u-v)$, and using the interval Poincar\'e inequality
in the width direction, also gives
\begin{align}
 \int_{R_\epsilon}|u|^2\,\dd A
 &\leq K\epsilon\ell(|p_1|^2+|p_2|^2)
 +K\ell^2\int_{R_\epsilon}|\partial_z u|^2\,\dd A
 +K\epsilon^2\int_{R_\epsilon}|\partial_y u|^2\,\dd A.
 \label{eq:neckmass}
\end{align}
These estimates retain possible neck-localized fields rather than
assuming they equilibrate instantaneously.

Choose a first positive transverse eigenfunction with
$\int u\,\dd A=0$ and $\int |u|^2\,\dd A=1$. The trial upper bound
\eqref{eq:transverseupper} gives total energy $E=O(\epsilon)$.
Equations~\eqref{eq:chamberpoincare}--\eqref{eq:porttrace} show
$u-U_i\to0$ in each chamber and $p_i-U_i=o(1)$.
Equation~\eqref{eq:neckmass} gives vanishing neck mass. Mean zero and
normalization then imply
\begin{equation}
 A_1U_1+A_2U_2=o(1),\qquad
 A_1U_1^2+A_2U_2^2=1+o(1),\qquad
 (U_1-U_2)^2=\frac1{A_1}+\frac1{A_2}+o(1).
 \label{eq:chambernormalization}
\end{equation}
The lower bound \eqref{eq:necklower} matches the trial upper bound,
and proves
\begin{equation}
 \widetilde\nu_1(\epsilon)
 =\frac\epsilon\ell\left(\frac1{A_1}+\frac1{A_2}\right)[1+o(1)].
 \label{eq:undilatedneckrate}
\end{equation}
Rescaling gives \eqref{eq:transverseasymptotic} and the chamber
limits \eqref{eq:chambercontrastshape}. This argument also explains
why chamber-access resistance does not change the leading prefactor
when $\ell$ is fixed and positive.

To prove a positive competing scale, suppose for contradiction that
$\widetilde\nu_2(\epsilon_k)\to0$ along a sequence. The first three
orthonormal eigenfunctions, including the constant, then all have
vanishing energy. Their chamber means remain bounded by normalization.
Although the unweighted port values need not stay bounded for such a
general low-energy sequence, \eqref{eq:porttrace} gives
\begin{equation}
 \epsilon_k|p_i|^2
 \leq K\epsilon_k|U_i|^2
 +K\epsilon_k[1+\ln(r_*/\epsilon_k)]E_i\longrightarrow0.
 \label{eq:weightedport}
\end{equation}
Equation~\eqref{eq:neckmass} therefore again forces the neck mass to
vanish. After taking subsequences, each eigenfunction converges in the
chambers to a vector of two constants. Their three pairwise inner
products converge to the weighted inner products in
$(\mathbb R^2,\operatorname{diag}(A_1,A_2))$. This would give three
orthonormal vectors in a two-dimensional space, a contradiction. Hence
$\liminf_{\epsilon\to0}\nu_2(\epsilon)>0$ independently of an
explicit formula for its limit.

Identifying that limit as \eqref{eq:transverselimit} additionally uses
the established convergence, with multiplicities, to the chamber
Neumann spectra and the collapsed-neck Dirichlet spectrum
\cite{Arrieta1995,JimboKosugi2009}. Neck-localized modes have amplitude of order
$\epsilon^{-1/2}$ when normalized in the shrinking neck; their finite
rates explain why the neck contributes $\pi^2/\ell^2$ to the positive
competing scale despite its vanishing area.

%
%

\ack{P. Z. acknowledges the support from the Adolf Martens Postdoctoral Fellowship (BAM-AMF-2025-1).}

\data{All data that support the findings of this study are included within the article (and any supplementary files).}

\end{document}